\documentclass[prc,showpacs,twocolumn,superscriptaddress]{revtex4}
\usepackage{graphicx}    

\begin{document}
\title{A Fully Integrated Transport Approach to Heavy Ion Reactions with an Intermediate Hydrodynamic Stage}

\author{Hannah Petersen}
\affiliation{Frankfurt Institute for Advanced Studies (FIAS),
Ruth-Moufang-Str.~1, D-60438 Frankfurt am Main,
Germany}
\affiliation{Institut f\"ur Theoretische Physik, Goethe-Universit\"at, Max-von-Laue-Str.~1, 
D-60438 Frankfurt am Main, Germany}
  
\author{Jan Steinheimer}
\affiliation{Institut f\"ur Theoretische Physik, Goethe-Universit\"at, Max-von-Laue-Str.~1,
 D-60438 Frankfurt am Main, Germany}
 
\author{Gerhard Burau}
\affiliation{Institut f\"ur Theoretische Physik, Goethe-Universit\"at, Max-von-Laue-Str.~1,
 D-60438 Frankfurt am Main, Germany}

\author{Marcus Bleicher}
\affiliation{Institut f\"ur Theoretische Physik, Goethe-Universit\"at, Max-von-Laue-Str.~1,
 D-60438 Frankfurt am Main, Germany}

\author{Horst St\"ocker}
\affiliation{Institut f\"ur Theoretische Physik, Goethe-Universit\"at, Max-von-Laue-Str.~1, 
D-60438 Frankfurt am Main, Germany} 
\affiliation{Gesellschaft f\"ur Schwerionenforschung (GSI), Planckstr.~1, D-64291 Darmstadt, Germany}
\affiliation{Frankfurt Institute for Advanced Studies (FIAS),
Ruth-Moufang-Str.~1, D-60438 Frankfurt am Main,
Germany}

\begin{abstract}
We present a coupled Boltzmann and hydrodynamics approach to relativistic heavy ion reactions. 
This hybrid approach is based on the Ultra-relativistic Quantum Molecular Dynamics (UrQMD) transport approach with an intermediate hydrodynamical evolution for the hot and dense stage of the collision. Event-by-event fluctuations are directly taken into account via the
non-equilibrium initial conditions generated by the initial collisions and string fragmentations in the microscopic UrQMD model. After a (3+1)-dimensional ideal hydrodynamic evolution, the hydrodynamical fields are mapped to hadrons via the Cooper-Frye equation and the subsequent hadronic cascade calculation within UrQMD proceeds to incorporate the important final state effects for a realistic freeze-out. This implementation allows to compare pure microscopic transport calculations with hydrodynamic calculations using exactly the same initial conditions and freeze-out procedure. The effects of the change in the underlying dynamics - ideal fluid dynamics vs. non-equilibrium transport theory - will be explored. The freeze-out and initial state parameter dependences are investigated for different observables. Furthermore, the time evolution of the baryon density and particle yields are discussed. We find that the final pion and proton multiplicities are lower in the hybrid model calculation due to the isentropic hydrodynamic expansion while the yields for strange particles are enhanced due to the local equilibrium in the hydrodynamic evolution. The results of the different calculations for the mean transverse mass excitation function, rapidity and transverse mass spectra for different particle species at three different beam energies are discussed in the context of the available data. 
\end{abstract}


\maketitle
\section{Introduction}
One of the main motivations to study high energy heavy ion collisions is the creation of a new deconfined phase of strongly interacting matter, the so called Quark-Gluon Plasma (QGP) \cite{Harris:1996zx,Bass:1998vz}. At the Relativistic Heavy Ion Collider (RHIC) many experimental observations like, e.g., jet quenching and high elliptic flow hint to the fact that a strongly coupled QGP (sQGP) might have been created \cite{Adams:2005dq,Back:2004je,Arsene:2004fa,Adcox:2004mh}. At CERN-SPS energies evidence for the creation of a new state of matter has been published, e.g., the enhanced K/$\pi$ ratio ('horn') and the step in the mean transverse mass excitation function for pions, kaons and protons \cite{:2007fe}. Especially the low energy (high $\mu_B$) program at SPS showed a culmination of exciting results. Therefore this energy regime will be the subject to further detailed studies at the CERN-SPS, BNL-RHIC, JINR-NICA and at the FAIR facility. 

Since the direct detection of free quarks and gluons is impossible due to the confining nature of QCD, it is important to model the dynamical evolution of heavy ion reactions to draw conclusions from the final state particle distributions about the interesting early stage of the reaction. One approach which aims at the description of heavy ion reactions consistently from the initial state to the final state is relativistic transport theory \cite{Bass:1998ca,Bleicher:1999xi,Molnar:2004yh,Xu:2004mz,Lin:2004en,Burau:2004ev}. This microscopic description has been applied quite successfully to the partonic as well as to the hadronic stage of the collision. Unfortunately, most transport approaches are restricted to $2 \rightarrow n$ scattering processes. Thus, if the particle density increases it becomes questionable if a restriction to two-particle interaction is still justified. While first attempts to include multi-particle interactions have been proposed \cite{Xu:2004mz,Barz:2000zz,Larionov:2007hy,Bleibel:2006xx,Bleibel:2007se}, this extension of transport theory is still in its infancy. To explain hadronization and the phase transition between the hadronic and the partonic phase on a microscopic level is also one of the main open issues that still has to be resolved. It is therefore difficult to find an appropriate prescription of the phase transition in such a microscopic approach. First, however promising attempts to solve the microscopic hadronization problem can be found in the literature \cite{Andersson:1977xs,Ellis:1995pe,Biro:1998dm,Traxler:1998bk,Hofmann:1999jx,Lin:2004en}. 

Hydrodynamics, on the other hand, has been proposed many years ago as a tool for the description of the hot and dense stage of heavy ion reactions where the matter might behave like a locally thermalized ideal fluid \cite{Hirano:2001eu,Kolb:2003dz,Nonaka:2006yn}. In this approach it is possible to model phase transitions explicitly because one of the major inputs to a hydrodynamic calculation is the equation of state (EoS). The hydrodynamic description has gained importance over the last few years because the high elliptic flow values that have been observed at RHIC seem compatible with some ideal hydrodynamic predictions \cite{Kolb:2000fha,Kolb:2001qz,Huovinen:2001cy}. The initial conditions and freeze-out prescription are the boundary conditions for a hydrodynamic calculation and therefore a further crucial input. Thus, the hydrodynamic results depend strongly on the initial and final state prescription that is applied in the specific calculation. 

To get a more consistent picture of the whole dynamics of heavy ion reactions various so called microscopic plus macroscopic (micro+macro) hybrid approaches have been launched during the last decade. Most noteworthy in this respect are the pioneering studies related to a coupling between UrQMD (Ultra-relativistic Quantum Molecular Dynamics) and hydrodynamics (a detailed systematic investigation of this coupling procedure can be found in the following references \cite{Dumitru:1999sf,Bass:1999tu,Bass:2000ib,Soff:2000eh,Soff:2001hc,Nonaka:2005aj,Nonaka:2006yn,Steinheimer:2007iy}).

Other approaches in the same spirit are, e.g., the NEXSpheRIO approach that uses initial conditions calculated in a non-equilibrium model (NEXUS) followed by an ideal hydrodynamic evolution \cite{Paiva:1996nv,Aguiar:2001ac,Socolowski:2004hw,Grassi:2005pm,Andrade:2005tx,Andrade:2006yh,Aguiar:2007zz} or a hybrid approach by Toneev et al. which uses QGSM initial conditions followed by a three-dimensional hydrodynamic evolution \cite{Skokov:2005ut}. In this way event-by-event fluctuations are taken into account and the calculation mimics more realistically the experimental case. For the freeze-out NEXspheRIO employs a continuous emission scenario or a standard Cooper-Frye calculation. Other groups, e.g., Teaney et al. \cite{Teaney:2001av}, Hirano et al. \cite{Hirano:2005xf,Hirano:2007ei}, Bass/Nonaka \cite{Nonaka:2006yn}, are using smooth Glauber or Color Glass Condensate initial conditions followed by a full two- or three-dimensional hydrodynamic evolution and calculate the freeze-out by a subsequent hadronic cascade. The separation of chemical and kinetic freeze-out and final state interactions like resonance decays and rescatterings are taken into account. There are two major conclusions from these previous studies: The treatment of the initial state fluctuations and the final decoupling is of major importance for a sound interpretation of the experimental data. 

Unfortunately, all presently existing micro+macro approaches rely on a complete separation of the three main ingredients (initial conditions, hydrodynamic evolution, transport calculation). Thus, it is impossible to compare the evolution of the system between hydrodynamics and transport simulation directly and from the same initial conditions. This may provide essential new insights into the role of viscosity and local equilibration. In addition, the usual separation of the program code does not allow for a dynamical coupling between hydrodynamics and transport calculation, which would be desirable to consistently solve the freeze-out puzzle \cite{Anderlik:1998et,Magas:1999yb,Bugaev:1999wz,Bugaev:2002ch}. 

To overcome these restrictions, we go forward and build a transport approach with an embedded three-dimensional ideal relativistic one fluid evolution for the hot and dense stage of the reaction. This allows to reduce the parameters for the initial conditions and the freeze-out prescription. The aim is to compare calculations with different EoS within the same framework. It will be possible to extract the effect of changes in the EoS, e.g., a phase transition from hadronic matter to the QGP, on observables. 

In this paper we describe the specific micro+macro hybrid approach that embeds a hydrodynamic phase in the UrQMD approach. First we explain the initial conditions, then introduce the basics of the hydrodynamic evolution including the hadron gas EoS and the transport calculation and illustrate how the freeze-out is treated. In the second part, the sensitivity of the results on the parameters are tested and the time evolution of the baryon density and the particle numbers are compared. Results on particle multiplicities and rapidity as well as transverse mass spectra are presented in the third part. 

At present we have calculated results imposing a hadron gas EoS to provide a baseline calculation to disentangle the effects of the different assumptions for the underlying dynamics in a transport vs. hydrodynamic calculation. The purely hadronic calculations can be compared in the broad energy regime from $E_{\rm lab}=2-160A~$GeV where a vast amount of experimental data from BNL-AGS and CERN-SPS exists and which will be explored in more detailed energy scans by the FAIR project near GSI and the critRHIC program. Studies employing different EoS are delayed to future work to concentrate on effects of the underlying dynamics first.           

\section{General aspects}
The modelling of the dynamical evolution of heavy ion reactions is essential to gain further insights about the properties of the newly produced hot and dense QCD matter. Transport theory aims at the description of all stages of the collision on the basis of an effective solution of the relativistic Boltzmann equation \cite{DeGroot:1980dk}  

\begin{equation}
\label{boltzmann}
p^\mu \cdot \partial_\mu f_i(x^\nu, p^\nu) = \mathcal{C}_i \quad .
\end{equation}

This equation describes the time evolution of the distribution functions for particle species $i$ and includes the full collision term on the right hand side. The interaction with external potentials leads to an additional term on the left hand side. The influence of potentials gets small at higher energies compared to the energy that is transferred by collisions. Therefore, they are dropped in Eqn. \ref{boltzmann} and are not further discussed here. Usually, the collision kernel is truncated on the level of binary collisions and $2\rightarrow n$ processes to keep the calculation numerically tractable. This microscopic approach has the advantage that it is applicable to non-equilibrium situations and the full phase space information is available at all stages of the heavy ion reaction. The restriction to binary collisions assumes large mean free paths of the particles. Between interactions the particle trajectories are given by straight line trajectories and particles are assumed to be in asymptotic states between the collisions (no ``memory effect'').

This assumption might not be justified in the hot and very dense stage of heavy ion collisions anymore. In this regime the continuum limit in form of relativistic hydrodynamics might fit better to the characteristics of the system. The hydrodynamic evolution is governed by the energy and momentum conservation laws for given initial conditions, i.e. spatial distributions of energy and net baryon number densities. The coordinate space is divided into small cells in which the distribution functions correspond to equilibrium distributions (Fermi or Bose distribution). In this macroscopic approach the propagated quantities are net baryon number and energy densities which can be translated into information about the temperature and chemical potential via the specific equation of state (EoS). Since the evolution is driven by pressure gradients and the pressure is determined via the EoS, the EoS is the essential ingredient for the hydrodynamical evolution. Thus, hydrodynamics is a good tool to describe collective behaviour. Ideal hydrodynamics applies to systems with small mean free path, otherwise viscous effects have to be taken into account \cite{Landau:1953gs}. A general advantage of hydrodynamics is the feature to explicitly incorporate phase transitions by changing the EoS.

However, in the late stage of the heavy ion reaction the system gets too dilute to apply ideal fluid dynamics. The hadronic rescatterings and decays of resonances have to be described, e.g., by using a transport description. Overall, there are two crucial points one has to take care of when building up a transport+hydrodynamics hybrid approach. The first is the initial switch from the microscopic to the macroscopic calculation where it has to be ensured that the local equilibrium assumption is fulfilled. The second one is the so called freeze-out where the hydrodynamic fields are mapped to particles that are further propagated in a hadronic cascade. The freeze-out transition must be placed in a region where both descriptions are valid at the same time,e.g., the phase transition region. In the following the specific implementation developed here will be discussed in more detail. 

\section{Specific micro+macro approach}
\subsection{UrQMD Approach}
For our investigation, the UrQMD model (v2.3)~\cite{Bass:1998ca,Bleicher:1999xi} is applied to heavy ion reactions from $E_{\rm lab}= 2-160A$~GeV. This non-equilibrium transport approach constitutes an effective solution of the relativistic Boltzmann equation (see Eqn. \ref{boltzmann}). The underlying degrees of freedom are hadrons and strings that are excited in high energetic binary collisions. Mean fields can in principle be taken into account in this framework, but the model is run in the so called cascade mode without inter-particle potentials. To omit the potentials is reasonable because the inclusion of mean fields would not change the results in the energy range that we are considering here. Note that this is consistent with the calculation of the equation of state for the hydrodynamic evolution where no mean field has been taken into account as well. 

The projectile and target nuclei are initialised according to a Woods-Saxon profile in coordinate space and Fermi momenta are assigned randomly for each nucleon in the rest frame of the corresponding nucleus. The hadrons are propagated on straight lines until the collision criterium is fulfilled. If the covariant relative distance $d_{\rm trans}$ between two particles gets smaller than a critical distance that is given by the corresponding total cross section a collision takes place,

\begin{equation}
d_{\rm trans}\le d_{0}=\sqrt{\frac{\sigma_{\rm tot}}{\pi}},\quad \sigma_{\rm tot}=\sigma(\sqrt{s},\mbox{type})\, .
\end{equation}

Each collision process is calculated in the rest frame of the binary collision. The reference frame that is used for the time ordering of the collisions and later on also for the switchings to and from the hydrodynamic phase is the equal speed-system of the nucleus-nucleus collision (for symmetric systems the equal speed system is identical to the center of mass system). 

In UrQMD 55 baryon and 32 meson species, ground state particles and all resonances with masses up to $2.25$ GeV, are implemented with their specific properties and interaction cross sections. In addition, full particle-antiparticle symmetry is applied. Isospin symmetry is assumed and only flavour-SU(3) states are taken into account. The elementary cross sections are calculated by detailed balance or the additive quark model or are fitted and parametrized according to the available experimental data. For resonance excitations and decays the Breit-Wigner formalism, utilizing their vacuum properties is employed.

Towards higher energies, the treatment of sub-hadronic degrees of freedom is of major importance. In the present model, these degrees of freedom enter via the introduction of a formation time for hadrons produced in the fragmentation of strings \cite{Andersson:1986gw,NilssonAlmqvist:1986rx,Sjostrand:1993yb}. String excitation and fragmentation is treated according to the Lund model. For hard collisions with large momentum transfer ($Q > 1.5$ GeV) PYTHIA is used for the calculation. A phase transition to a quark-gluon state is not incorporated explicitly into the model dynamics. However, a detailed analysis of the model in equilibrium yields an effective equation of state of Hagedorn type \cite{Bass:1997xw,Bravina:1998it,Belkacem:1998gy,Bravina:1999dh,Bravina:2008ra}. The UrQMD transport model is successful in describing the yields and the $p_{t}$ spectra of various particles in proton-proton, proton-nucleus and nucleus-nucleus collisions \cite{Bratkovskaya:2004kv}. A compilation of results of the actual version UrQMD-2.3 compared to experimental data can be found in \cite{Petersen:2008kb}. 

Apart from the success of transport simulations to describe spectra and yields certain problems remain: 
\begin{itemize}
\item Elliptic flow values above SPS energies are too small \cite{Bleicher:2000sx,Petersen:2006vm},
\item HBT radii hint to a very small $R_o/R_s$ ratio \cite{Kniege:2006in,Li:2007yd},
\item Strangeness, especially multi-strange baryons are not produced in sufficient amounts \cite{Soff:1999et}.
\end{itemize}

These observables that are sensitive to the early stage of the collision (pressure) or to the approach of thermal and chemical equilibrium during the collision history hint to the fact that a purely hadronic transport model may not be sufficient to describe the dynamics of the hot and dense stage of heavy ion reactions at higher energies \cite{Bleicher:2000sx,Molnar:2002bz,Heinz:2002un,Petersen:2006vm,Li:2007yd}. Therefore, these observations exemplify the need to embed a full three-dimensional relativistic fluid dynamics description for these stages of the reaction. 

For the results that are shown in this paper, the reference calculations are always performed employing the state-of-the-art UrQMD-2.3 model. 

\subsection{Initial Conditions}

The Ultra-relativistic Quantum Molecular Dynamics Model is used to calculate the initial state of a heavy ion collision for the hydrodynamical evolution \cite{Steinheimer:2007iy}. This is necessary to account for the non-equilibrium nature of the very early stage of the collision. Event-by-event fluctuations of the initial state are naturally included by this set-up. The coupling between the UrQMD initial state and the hydrodynamical evolution takes place when the two Lorentz-contracted nuclei have passed through each other. The initial time to begin with the hydrodynamical evolution is calculated via Eqn. \ref{eqn_tstart} (and is assumed to be at least 1 fm/c):

\begin{equation}
t_{start}=\frac{2 R}{\gamma v}=\frac{2 R}{\sqrt{\gamma^2-1}}=2 R \sqrt{\frac{2 m_N}{E_{\rm lab}}}\, ,
\label{eqn_tstart}
\end{equation}

where R is the radius of the nucleus, $m_N$ is the nucleon mass and $E_{\rm lab}$ is the kinetic beam energy. This assures that (essentially) all initial baryon-baryon scatterings have proceeded and that the energy deposition has taken place. This is the  earliest possible transition time where thermalization might be achieved \cite{Bravina:2008ra}. It is also convenient from the hydrodynamical point of view since at that time the two baryon currents that fly into oppposite directions have separated again.

In general, it is not well-established how and when chemical and kinetic equilibrium might have been reached in the early stage of the collision. One of the problems is, e.g., that the local equilibrium assumption might not apply equally well to all parts of the system at the same time in the computational frame which corresponds to the center of mass system of the two colliding nuclei. As a consequence, the faster particles have had less time in their local rest frame to equilibrate. For the bulk part and the high density region at midrapidity the difference between the two frames is small. These problems are present in all hydrodynamic/macroscopic approaches that rely on an equilibrium assumption and it is not our attempt to resolve these difficulties in this paper. One perspective might be the dynamical coupling between the initial transport calculation and the hydrodynamic evolution including source terms on both sides of the transition surface.
  
To allow for a consistent and numerically stable mapping of the 'point like' particles from UrQMD to the 3-dimensional spatial-grid with a cell size of $(0.2 \, {\rm fm})^3$, each hadron is represented by a Gaussian with a finite width. ``Pre-formed'' hadrons in the process of string fragmentation are also included in the transformation to the hydrodynamic quantities. I.e. each particle is described by a three-dimensional Gaussian distribution of its total energy, momentum (in x-, y-, and z-direction) and baryon number density. The width of these Gaussians is chosen to be $\sigma=1~$fm. A smaller Gaussian widths leads to numerical instabilities (e.g., entropy production) in the further hydrodynamical evolution, while a broader width would smear out the initial fluctuations to a large extent. To account for the Lorentz-contraction of the nuclei in the longitudinal direction, a Lorentz-gamma-factor is included. The resulting distribution function in the computational frame (cf), e.g., for the energy density, reads:

\begin{equation}
\epsilon_{\rm cf}(x,y,z)=N  e^{-\frac{(x-x_{p})^2+(y-y_{p})^2
+(\gamma_z(z-z_{p}))^2}{2 \sigma^2}}\quad,
\end{equation}

\noindent where $N=(1/2 \pi )^{3/2} \gamma_z/\sigma^3  E_{\rm cf} $ provides the proper normalisation, $\epsilon_{\rm cf} $ and $E_{\rm cf} $ are the energy density and total energy of the particle in the computational frame, while $(x_p,y_p,z_p)$ is the position vector of the particle. Summing over all single particle distribution functions leads to distributions of energy, momentum and baryon number densities in each cell.

To allow for calculations at finite impact parameter the spectators - nucleons that have not interacted at all before the start time of the hydrodynamic evolution $t_{start}$ - are propagated separately from the hydrodynamic evolution. The spectators are propagated on straight line trajectories in the usual cascade mode until the end of the hydrodynamic phase has been reached. 
 
Instead of smearing out the initial distributions by describing the point like hadrons as Gaussian distributions, one could also obtain a smooth distribution by averaging over a large sample of UrQMD events. Our procedure of creating a new initial state for each event is motivated by the fact, that the experimental results all relate to observed (averaged) final, and not initial, states. Thus, event-by-event fluctuations of the initial state can be observed, (e.g., in $v_2$ fluctuations) and have therefore been taken into account properly (for discussion of the importance of these fluctuations see, e.g., \cite{Paiva:1996nv,Drescher:2006ca}). 

\begin{figure}[h]
\includegraphics[width=0.5\textwidth]{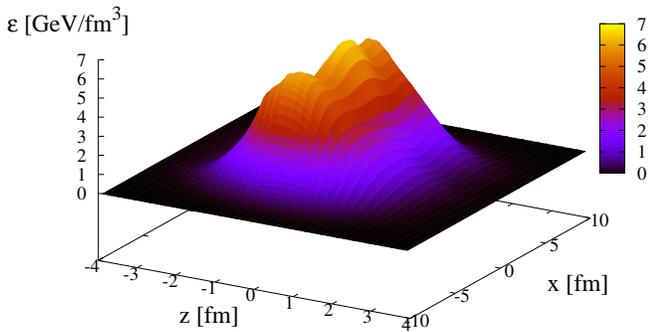}
\caption{\label{fig_epsilon_plane}
(Color online) Initial energy density distribution in the reaction plane ($x-z$ plane) of one central ($b=0$ fm) Pb+Pb collision at $E_{\rm lab}=40A~$GeV. $z$ corresponds to the beam direction and $x$ to the in-plane axis (direction of the impact parameter) of the collision.}
\end{figure}

\begin{figure}[h]
\includegraphics[width=0.5\textwidth]{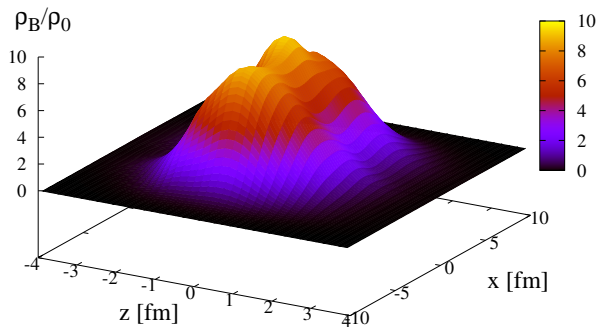}
\caption{\label{fig_rhob_plane}
(Color online) Initial net baryon number density distribution in the reaction plane ($x-z$ plane) of one central ($b=0$ fm) Pb+Pb collision at $E_{\rm lab}=40A~$GeV. $z$ corresponds to the beam direction and $x$ to the in-plane axis (direction of the impact parameter) of the collision.}
\end{figure}

As an example, Figs. \ref{fig_epsilon_plane} and \ref{fig_rhob_plane} show the energy and baryon number densities obtained in one single central ($b=0$ fm) Pb+Pb collision at $E_{\rm lab}=40A~$ GeV after the initialisation of the hydrodynamic fields. The starting time is in this case $t_{start}=2.83$ fm and the densities in the figures correspond to the same time. All quantities are given in the local rest frame. The maximum values reach 6 GeV/fm$^3$ for the energy density and around 8 times the nuclear ground state density for the baryon number density. The distributions are quite smooth which is necessary to provide possible initial conditions for the hydrodynamic evolution. One can see some peaks that correspond to local maxima of the distributions (``hot spots''). It is further clear that the single event distributions are not symmetric, neither in the transverse nor the longitudinal direction.  

The remaining question is if the system is thermalized enough to assure that the local equilibrium assumption of ideal hydrodynamics is fulfilled. In our case, the hydrodynamic code transforms all the given quantities from the computational frame to the local rest frame of the energy momentum tensor which is also known as the Landau frame. This frame coincides in ideal hydrodynamics with the Eckart frame which is defined as the local rest frame of the baryon number current. The iterative calculation of the cell velocity succeeds if those two frames are close enough to each other. By this transformation the system is forced to local equilibrium.    

\subsection{Hydrodynamic Evolution}
Ideal relativistic one fluid dynamics is based on the conservation of energy, momentum and the net baryon number current. For the hydrodynamical evolution local equilibrium is assumed and zero viscosity which corresponds to zero mean free path. The two conservation equations that govern the evolution are \cite{Landau:1953gs,Clare:1986qj}

\begin{equation}
\partial_\mu T^{\mu\nu}=0 \quad \mbox{and } \quad \partial_\mu N^\mu=0,
\end{equation}
where $T^{\mu\nu}$ is the energy-momentum tensor and $N^\mu$ is the baryon current. 
For an ideal fluid the energy-momentum tensor and the net baryon number current take the simple form

\begin{equation}
T^{\mu\nu}=(\epsilon_{\rm lrf}+P)\,u^\mu \,u^\nu -P \,g^{\mu\nu}\quad \mbox{and } \quad N^\mu=\rho_{\rm lrf}\, u^\mu
\end{equation}
where $\epsilon_{\rm lrf}, P$ and $\rho_{\rm lrf}$ are the local rest frame energy density, pressure and net baryon density, respectively. $u^\mu=\gamma (1,\vec{v})$ is the four velocity of the cell and $g^{\mu\nu}=diag(+,-,-,-)$ is the metric tensor. The local rest frame is defined as the frame where $T^{\mu\nu}$ has diagonal form, (i.e. all off-diagonal elements vanish). The four-velocity of the cells is calculated via the transformation into the local rest frame.  

The equations of motion are solved in the following form by employing computational frame quantities $\epsilon_{\rm cf}, p^i$ and $\rho_{\rm cf}$ for the energy, momentum and net baryon number densities. 
\begin{eqnarray}
\partial_t \epsilon_{\rm cf}+\nabla \cdot(\epsilon_{\rm cf}\,\vec{v})&=&-\nabla\cdot(P\,\vec{v})\\
\partial_t \vec{p}+\nabla \cdot(\vec{p}\,\vec{v})&=&-\nabla P\\
\partial_t \rho_{\rm cf}+\nabla \cdot(\rho_{\rm cf}\,\vec{v})&=&0
\end{eqnarray}
 
In our case, the full (3+1) dimensional hydrodynamic evolution is performed using the SHASTA algorithm \cite{Rischke:1995ir,Rischke:1995mt}. The partial differential equations are solved on a three-dimensional spatial Eulerian grid with fixed position and size in the computational frame. The standard size of the grid is 200 cells in each direction while the cell size has been chosen to be $(dx)^3=(0.2\, \rm{fm})^3$ which leads to timesteps of $dt=0.08$ fm. Depending on the beam energy, the cell sizes may require adjustment to assure a stable solution of the differential equation.  

The equation of state is needed as an additional input to calculate the pressure, temperature and chemical potential corresponding to the energy and the baryon number densities. Since the evolution of the system is driven by pressure gradients the EoS has the most important influence on the evolution.  

\subsection{Equation of State}

To solve the hydrodynamical equations, the EoS, the pressure as a function of energy and net-baryon number density, is needed as an input. Since the actual EoS of hot and dense QCD matter is still not precisely known, it may seem disadvantageous to have this additional uncertainty in the model. On the contrary it may prove to be an important trait of the model to be able to study changes on the dynamics of the bulk matter when changing the EoS thus finding observables for a phase transition in hot QCD matter. For recent discussions of different EoS and how to obtain EoS from lattice calculations, the reader is referred to \cite{Bluhm:2007nu,Redlich:2007ud,Biro:2006zy}. 
\\
The EoS used in the present calculations is a grand canonical description of a free, non interacting gas of hadrons. We will refer to it as the Hadron Gas (HG).
It follows from the hadronic chiral model presented in \cite{Papazoglou:1998vr,Zschiesche:2006rf}. The chiral hadronic $SU(3)$ Lagrangian incorporates the complete set of
baryons from the lowest flavour-$SU(3)$ octet, as well as
the entire multiplets of scalar, pseudo-scalar, vector and axial-vector
mesons \cite{Papazoglou:1998vr}. In mean-field approximation, the
expectation values of the scalar fields relevant for symmetric nuclear
matter correspond to the non-strange and strange chiral quark
condensates, namely the $\sigma$ and its $s\bar{s}$ counterpart
$\zeta$, respectively, and further the $\omega$ and $\phi$ vector
meson fields. Another scalar iso-scalar field, the dilaton $\chi$, is
introduced to model the QCD scale anomaly. However, if $\chi$
does not couple strongly to baryonic degrees of freedom then it remains
essentially ``frozen'' below the chiral transition \cite{Papazoglou:1998vr}.
Consequently, we focus here on the role of the
quark condensates.

Interactions between baryons and scalar (BM) or vector (BV)
mesons, respectively, are introduced as
\begin{eqnarray}
\label{L_BM+V}
{\cal L}_{\rm BM} &=&
-\sum_{i}   \overline{\psi}_i \left( g_{i\sigma}\sigma + g_{i\zeta}\zeta
\right) \psi_i
~,\\
{\cal L}_{\rm BV}
  &=& -\sum_{i}   \overline{\psi}_i
\left( g_{i\omega}\gamma_0\omega^0 + g_{i \phi}\gamma_0 \phi^0 \right) \psi_i
~,
\end{eqnarray}
Here, $i$ sums over the baryon octet ($N$, $\Lambda$, $\Sigma$, $\Xi$).
A term ${\cal L}_{\rm vec}$ with
mass terms and quartic self-interaction of the vector mesons is also added:
\begin{eqnarray}
{\cal L}_{\rm vec} &=& \frac{1}{2}
a_\omega \chi^2 \omega^2 + \frac{1}{2}
a_\phi \chi^2 \phi^2
+ g_4^{\,4} (\omega^4 + 2 \phi^4 )~. \nonumber
\end{eqnarray}
The scalar self-interactions are
\begin{eqnarray}
\label{L_0}
{\cal L}_0 &=& -\frac{1}{2} k_0 \chi^2 (\sigma^2+\zeta^2) + k_1
    (\sigma^2+\zeta^2)^2 + k_2 ( \frac{ \sigma^4}{2} + \zeta^4)
    \nonumber \\ & +& k_3 \chi \sigma^2 \zeta
 - k_4 \chi^4 - \frac{1}{4}\chi^4  \ln\frac{\chi^4}{\chi_0^{\,4}}
   +\frac{\delta}{3} \chi^4 \ln\frac{\sigma^2\zeta}{\sigma_0^{\,2} \zeta_0}
~.
\end{eqnarray}
Interactions between the scalar mesons induce the spontaneous
breaking of chiral symmetry (first line) and the scale breaking via
the dilaton field $\chi$ (last two terms).

Non-zero current quark masses break chiral symmetry explicitly in
QCD. In the effective Lagrangian this corresponds to terms such as
\begin{eqnarray}
{\cal L}_{\rm SB} &=& -\frac{\chi^2}{\chi_0^{\,2}}
\left[m_\pi^2 f_\pi \sigma + (\sqrt{2}m_K^2 f_K -
\frac{1}{\sqrt{2}}
m_{\pi}^2 f_{\pi})\zeta \right]~.
\end{eqnarray}

According to ${\cal L}_{\rm BM}$ (\ref{L_BM+V}), the effective
masses of the baryons,
$m_i^*(\sigma,\zeta)=g_{i\sigma}\,\sigma+g_{i\zeta}\,\zeta$\,,
are generated through their coupling to the chiral condensates,
which attain non-zero vacuum expectation values due to their
self-interactions \cite{Papazoglou:1998vr} in ${\cal L}_0$ (\ref{L_0}).
The effective masses of the mesons are obtained as the second
derivatives of the mesonic potential
${\cal V}_{\rm Meson}
 \equiv
 - {\cal L}_0 - {\cal L}_{\rm vec} - {\cal L}_{\rm SB}
$
about its minimum.\\
\\
All parameters of the chiral model discussed so far are fixed by either
symmetry relations, hadronic vacuum observables or nuclear matter
saturation properties (for details see \cite{Papazoglou:1998vr}).
In addition, the model also provides a satisfactory description of realistic
(finite-size and isospin asymmetric) nuclei and of neutron stars
\cite{Papazoglou:1998vr,Schramm:2002xa,Schramm:2002xi}.
%

%\subsection{Heavy baryonic degrees of freedom}
If the baryonic degrees of freedom are restricted to the members
of the lowest lying octet,
the model exhibits a smooth decrease of the chiral condensates
(crossover) for both high $T$ and high $\mu_B$  \cite{Papazoglou:1998vr,Zschiesche:2001dx}.
However, additional baryonic degrees of freedom changes this into a first-order phase transition
in certain regimes of the $T$-$\mu_B$ plane, depending on the
couplings \cite{Theis:1984qc,Zschiesche:2001dx,Zschiesche:2004si,Zschiesche:2006rf,Steinheimer:2007iy}.\\
\\
In what follows, the meson fields are replaced by their (classical)
expectation values, which corresponds to neglecting quantum and
thermal fluctuations. Fermions have to be integrated out
to one-loop. The grand canonical potential can then be written as
\begin{eqnarray}
\label{thermpot}
   {\Omega}/{V}&=& -{\cal L}_{\rm vec} - {\cal L}_0 - {\cal L}_{\rm SB}
-{\cal V}_{\rm vac} \\
& &{} -T \sum_{i \in B} \frac{\gamma_i }{(2 \pi)^3} %B=N,\Lambda,\Sigma,\Xi,R
\int d^3k \left[\ln{\left(1 + e^{-\frac 1T[E^{\ast}_i(k)-\mu^{\ast}_i]}\right)} \right]
\nonumber \\
& &{}+T \sum_{l\in M} \frac{\gamma_l}{(2 \pi)^3} %M=\pi,K,\eta,\omega,\rho,\phi,k^\ast
\int d^3k \left[\ln{\left(1 - e^{-\frac 1T[E^{\ast}_l(k)-\mu^{\ast}_l]}\right)
}\right], \nonumber
\end{eqnarray}
where $\gamma_B, \gamma_M$ denote the baryonic and mesonic
spin-isospin degeneracy factors and $E^{\ast}_{B,M} (k)
= \sqrt{{k}^2+{m_{B,M}^*}^2}$ are the corresponding single
particle energies. The effective baryon-chemical potentials are $\mu^{\ast}_i = \mu_i-g_{i \omega} \omega-g_{i \phi} \phi$, with
$\mu_i= (n^i_q - n^i_{\bar{q}}) \mu_q + (n^i_s - n^i_{\bar{s}})
\mu_s$. Here $\mu_q = \mu_B / 3$ is the quark, and $\mu_s$ the strange quark, chemical potential. The potentials of the mesons are given by the sum of the corresponding quark and anti-quark chemical potentials.  The
vacuum energy ${\cal V}_{\rm vac}$ (the potential at $\rho_B=T=0$) has
been subtracted.

By extremizing $\Omega/V$ one obtains self-consistent gap equations
for the meson fields. Here, globally non-strange matter is considered and 
$\mu_S$ for any given $T$ and $\mu_B$ is adjusted to obtain a vanishing net strangeness.\\
\\
Once the grand canonic potential is known as a function of $T$ and $\mu_B$, all other thermodynamic quantities are derived straightforward. In its minima the grand canonic potential corresponds to $-p$, the pressure. The entropy density $s$, number density $n$ and energy density $\epsilon$ then follow from the Euler relation,
\begin{equation}
	\epsilon = -p + s T + \sum_{i}{\mu_i n_i}
\end{equation}
where the sum runs over all included hadron species.\\
\\
Setting all hadron masses and chemical potentials to their vacuum values, and adding all reliably known heavy resonance states - with masses up to 2 GeV \cite{Eidelman:2004wy} - as free particles into (\ref{thermpot}), yields the above mentioned Hadron Gas EoS \cite{Zschiesche:2002zr}.
Hence, the hadronic degrees of freedom included in this EoS are consistent with the active degrees of freedom in the UrQMD model. This enables us to directly compare the dynamics of the hydrodynamic model with the transport simulation.\\
\\
Using the chiral model and adding additional baryonic degrees of freedom as well as adjusting their scalar and vector coupling, an EoS with a phase structure including a first order phase transition and even a critical endpoint at finite $\mu_B$ can be obtained \cite{Zschiesche:2006rf}. This chiral EoS has already been successfully applied to a hydrodynamic calculation \cite{Steinheimer:2007iy}. Here the essentially different equation of state leads to distinguishable different results on the properties of bulk matter.\\
To emphasize these differences even more, a MIT-Bag model EoS matched to an interacting hadron gas \cite{Rischke:1995mt}, generating a phase structure with a broad first order phase transition at all $\mu_B$, can also be applied in our model.\\
Consequently these results will be compared to our present calculations, constituting observables for a phase transition in hot and dense QCD matter, even suggesting the order of this phase transition. Comparisons between the different EoS (i.e. free Hadron Gas, chiral EoS, and Bag model EoS) will be presented in a follow-up paper.

\subsection{Freeze-out}
\label{freeze-out}
\begin{figure}[h]
\includegraphics[width=0.5\textwidth]{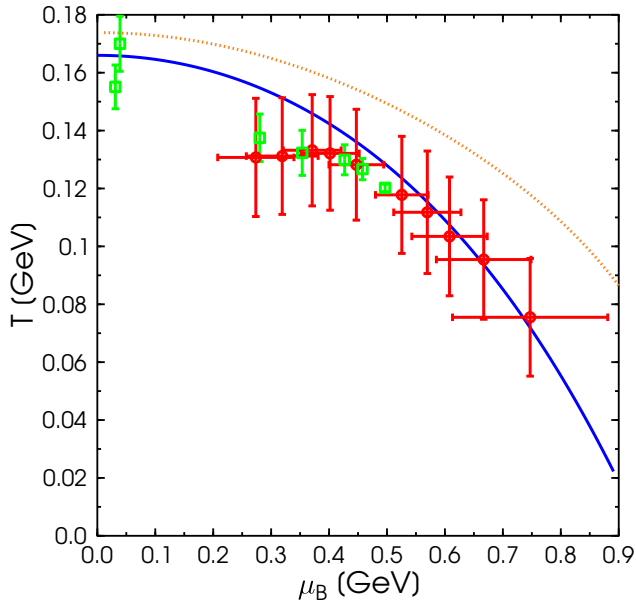}
\caption{\label{fig_tmu_width}
(Color online) Mean values of temperatures and baryo-chemical potentials at freeze-out for different beam energies are depicted as red circles (starting in the lower right corner at $E_{\rm lab}=2A~$GeV and going through 4,6,8,11,20,30,40 and 80 to $E_{\rm lab}=160A~$GeV in the upper left). The error bars indicate the width of the distribution. The dotted line depicts the line of constant energy density ($\epsilon =5 \cdot \epsilon_0$) that corresponds to our freeze-out criterium. For comparison the freeze-out line calculated by Cleymans et al. \cite{Cleymans:1998fq,Cleymans:2006qe} (full line) and results from Dumitru et al. \cite{Dumitru:2005hr}(green open squares) are shown.}
\end{figure}

Presently, the hydrodynamic evolution is stopped, if the energy density drops below five times the nuclear ground state energy density (i.e. $\sim 730$ MeV/${\rm fm}^3$) in all cells. This criterium corresponds to a T-$\mu_B$-configuration where the phase transition is expected (see dotted line in Fig. \ref{fig_tmu_width}), i.e. a region where the hydrodynamic and the transport description are valid at the same time. The hydrodynamic fields are mapped to hadrons according to the Cooper-Frye equation \cite{Cooper:1974mv} 

\begin{equation}
\label{cooper_frye}
E \frac{dN}{d^3p}=\int_\sigma f(x,p) p^\mu d\sigma_\mu
\end{equation}

where $f(x,p)$ are the boosted Fermi or Bose distributions corresponding to the respective particle species. Since we are dealing with an isochronous freeze-out, the normal vector on the hypersurface is $d\sigma_\mu = (d^3x,\vec{0})$. 

Let us note that it is of utmost importance to consider the same degrees of freedom on both sides of the hypersurface because otherwise energy and momentum conservation is violated. In our case, this is assured by the inclusion of the same particle species in the equation of state for the hydrodynamic calculation as in the transport calculation. In principle, it might also happen that particles are moving back into the hydrodynamic phase, however, the explosive character of heavy ion reactions, i.e. the rapid expansion flow suppresses the back-streaming effect. Therefore, this effect is negligible in our situation \cite{Bugaev:2004kq}.  

The assumption of an isochronous freeze-out leads to fluctuations of the  temperature and baryo-chemical potential distributions and not to single values for some of the thermodynamic quantities on the hypersurface. The quality of this approach and that the two (hydrodynamics and transport) prescriptions are valid through the applied range of switching temperatures is shown in the next section, where parameter tests and the time evolutions of the particle yields are explored in detail. 

\begin{figure}[h]
\includegraphics[width=0.5\textwidth]{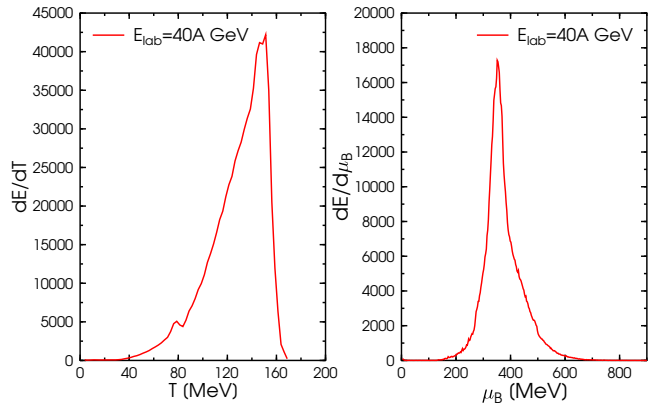}
\caption{\label{fig_dedtempdmu}
(Color online) Distribution of the energy in the cells at freeze-out at $E_{\rm lab}=40A~$GeV.}
\end{figure}

Fig. \ref{fig_dedtempdmu} shows the distribution of energy in the cells at freeze-out with respect to temperature and baryo-chemical potential at $E_{\rm lab}=40A~$GeV. We present here the energy distribution and not the number of cells because it is more interesting to know where the energy of the system sits than considering the many almost empty cells that do essentially not contribute to particle production. From Fig. \ref{fig_dedtempdmu} one obtains the mean values of the distributions that are in line with results from statistical model fits. The mean values of, e.g., the temperature can be calculated as

\begin{equation}
\langle T \rangle = \frac{\sum_{i,j,k} \,T(i,j,k) \rho_B(i,j,k)}{\sum_{i,j,k}\rho_B(i,j,k)},
\end{equation}

where $i,j,k$ are the cell indices and the sum runs over all cells of one event. The net baryon number density $\rho_B$ has been used as a weighting factor.  

To illuminate this finding in more detail, Fig. \ref{fig_tmu_width} shows the mean values of the temperature and baryo-chemical potential distributions at different energies in the $T-\mu_B$-plane for central ($b=0$ fm) Au+Au/Pb+Pb collisions. Also, the widths of the distributions are depicted as ``error'' bars. Fig. \ref{fig_tmu_width} shows that the present freeze-out distributions are similar to the parametrized curve for chemical freeze-out as calculated by Cleymans et al. from statistical model fits to final particle multiplicities. The calculation by Dumitru et al. shows mean values as well as widths of temperature and baryo-chemical potential distributions that have been obtained by statistical model fits to final particle yields employing the assumptions of an inhomogeneous freeze-out hypersurface. This calculation also leads to similar values as our calculation. 

\begin{figure}[h]
\includegraphics[width=0.5\textwidth]{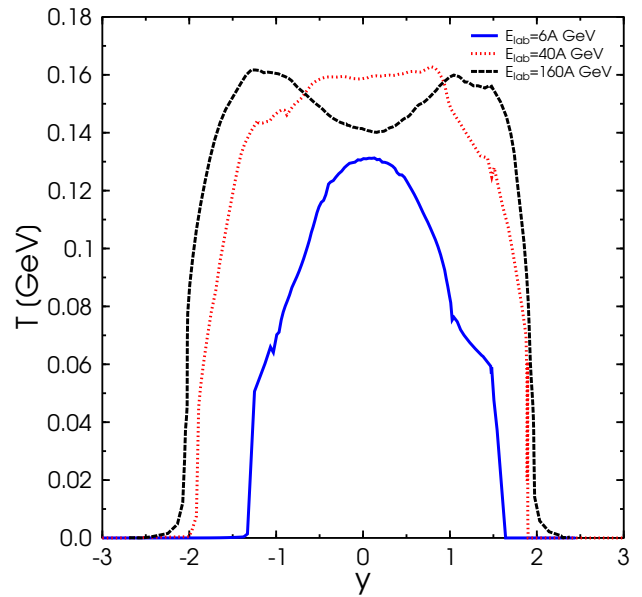}
\caption{\label{fig_temprap}
(Color online) Rapidity profile of the freeze-out temperatures in the spatial plane with $x=y=0$ fm for central Au+Au/Pb+Pb collisions at three different beam energies ($E_{\rm lab}=6,~40$ and $160A~$GeV).}
\end{figure}

The effect that the mean temperature at the transition to the transport prescription saturates or even drops down a little at higher beam energies is related to the rapidity distribution of the temperature in the hydrodynamic cells at freeze-out which is shown in Fig. \ref{fig_temprap} for three different beam energies. At low beam energies the midrapidity region coincides with the hottest region at freeze-out. At higher SPS energies the situation changes. The hottest cells are at high rapidities while the midrapidity region has already cooled down well below the temperature of 170 MeV. This problem might be resolved by a different freeze-out prescription on another hypersurface (e.g.,isotherm, iso-$\epsilon$) and is subject to future investigations. The best solution will be the dynamical coupling between hydrodynamics and transport which allows also for back-streaming contributions.   

In the following the practical implementation will be explained in more detail. The implementation is based on a Monte Carlo sampling of Eqn. \ref{cooper_frye} and follows the general steps: 

\begin{enumerate}
\item
The particle numbers $N_i$ are calculated according to the following formula,
\begin{equation} 
N_i= n_i \cdot \gamma \cdot V_{cell}=\int d^3p f_i(x,p) \cdot \gamma \cdot V_{cell}
\end{equation}
where the index $i$ runs over the different particle species like, e.g., $\pi$, p, $\rho$ or $\Delta$. $\gamma$ is the boost factor between the computational frame and the cell. $V_{cell}$ is the volume of the cell in the computational frame and $n$ is the particle number density. All cells with temperatures that are lower than 3 MeV are discarded from the following procedure because of numerical reasons. The local rest frame equilibrium distribution function is denoted by $f_i(x,p)$. To simplify the calculation, a relativistic Boltzmann distribution is used for all particles, except pions. It has been checked, that the Boltzmann approximation is sufficient to describe all particle species it is applied to. For the Boltzmann distribution the momentum integration leads to the following result for the particle number density 

\begin{equation}
n_i=\frac{4\pi g m^2 T}{(2\pi)^3}\exp{\left(\frac{\mu}{T}\right)}K_2\left(\frac{m}{T}\right)
\end{equation}

where $g$ is the degeneracy factor for the respective particle species, $m$ is the mass of the particle to be produced, $T$ the temperature of the cell and $K_2$ is the modified Bessel function. The chemical potential $\mu$ includes the baryo-chemical potential and the strangeness chemical potential in the following way

\begin{equation}
\mu=B\cdot \mu_B+S\cdot \mu_S
\end{equation}  

where $S$ is the quantum number for strangeness and $B$ is the baryon number.
 
For pions the Bose distribution has to be taken into account because the pion mass is on the order of the temperature of the system. In this case, the momentum integration involves an infinite sum over modified Bessel functions

\begin{equation}
n_\pi =\frac{g_\pi m_\pi^2T}{(2\pi)^2}\sum_{k=1}^\infty \frac{1}{k}K_2\left(\frac{k m_\pi}{T}\right) \, .
\end{equation} 

To calculate the number of particles in the computational frame the particle number density has to be multiplied with the Lorentz-stretched volume of the cell ($V_{\rm cell}=(0.2)^3 {\rm fm}^3$).

\item   
The average total number of particles in the cell, $\left\langle N \right\rangle$, is the sum over all particle numbers $N_i=n_i \gamma V_{\rm cell}$

\begin{equation}
\langle N \rangle =\sum_i N_i\,.
\end{equation}

\item
The total number of particles emitted from a cell, $N_i$, is obtained from a Poisson distribution according to $P(N)= \frac{\left\langle N \right\rangle^{N}}{N!}e^{-\left\langle N \right\rangle}$.

In the limit of small mean values, the Poisson distribution becomes $P(1)\approx \left\langle N \right\rangle$. Thus it can be decided by one random number between 0 and 1 if a particle is produced in the respective cell. If the random number is smaller than $\left\langle N \right\rangle$ one particle is produced and there is no particle production otherwise. The full Poisson distribution is used, if the particle number $\left\langle N \right\rangle$ is larger than ($0.01$). This assures an accuracy better than 1 \%.

\item
The particle type is chosen according to the probabilities $N_i/\langle N \rangle$. 

\item
The $I_3$ component of the isospin is distributed randomly because UrQMD assumes full isospin symmetry. To conserve the overall charge of the system and the initial isospin-asymmetry the probability to generate the isospin component that leads to the right value of the charge that should be obtained in the end is favoured. The other isospin components are exponentially suppressed. The power of the exponential is proportional to the difference of the  total charge generated by this produced particle and the required value. 
 
\begin{figure}[h]
\includegraphics[width=0.5\textwidth]{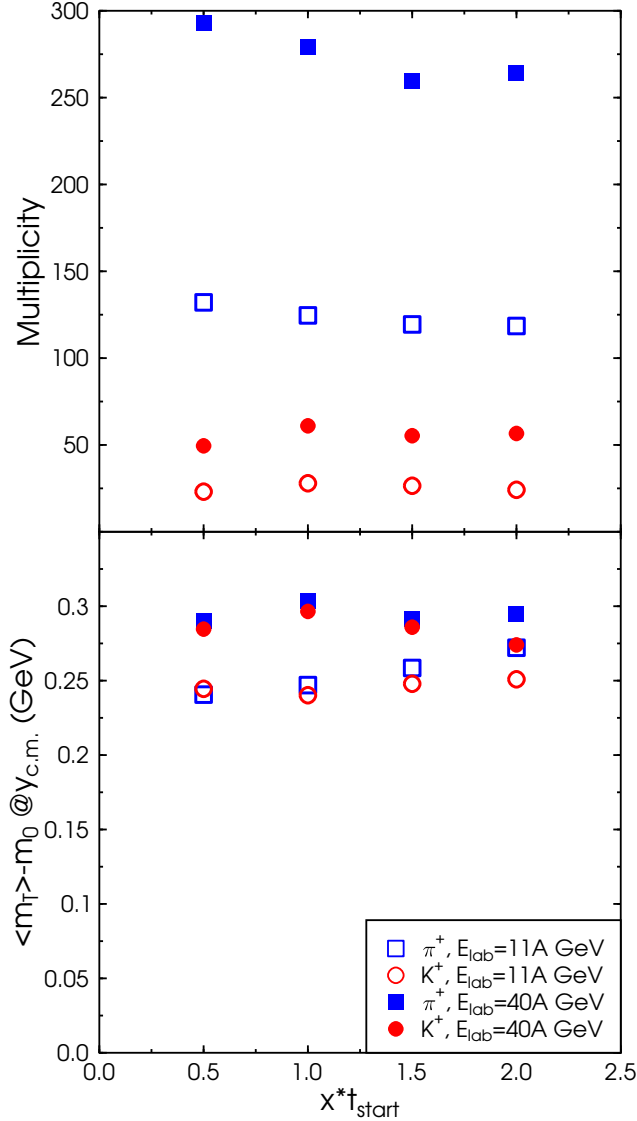}
\caption{\label{fig_intestmul}
(Color online) Pion and kaon multiplicities (upper panel) and mean transverse mass of pions and kaons at midrapidity ($|y|<0.5$) (lower panel) for four different starting times for central ($b<3.4$ fm) Au+Au/Pb+Pb collisions. The open symbols depict the result at $E_{\rm lab}=11A~$GeV while the filled symbols are the results at $E_{\rm lab}=40A~$GeV.}
\end{figure}

\item 
The 4-momenta of the particles are generated according to the Cooper-Frye equation (see Eqn. \ref{cooper_frye}). For baryons and strange mesons the chemical potentials for baryon number and strangeness are taken into account.
 
\item
The particle vector information is transferred back into the UrQMD model. The subsequent hadronic cascade calculation incorporates important final state effects as, e.g., rescatterings of the particles and resonance decays. 
\end{enumerate}

\begin{figure}[h]
\includegraphics[width=0.5\textwidth]{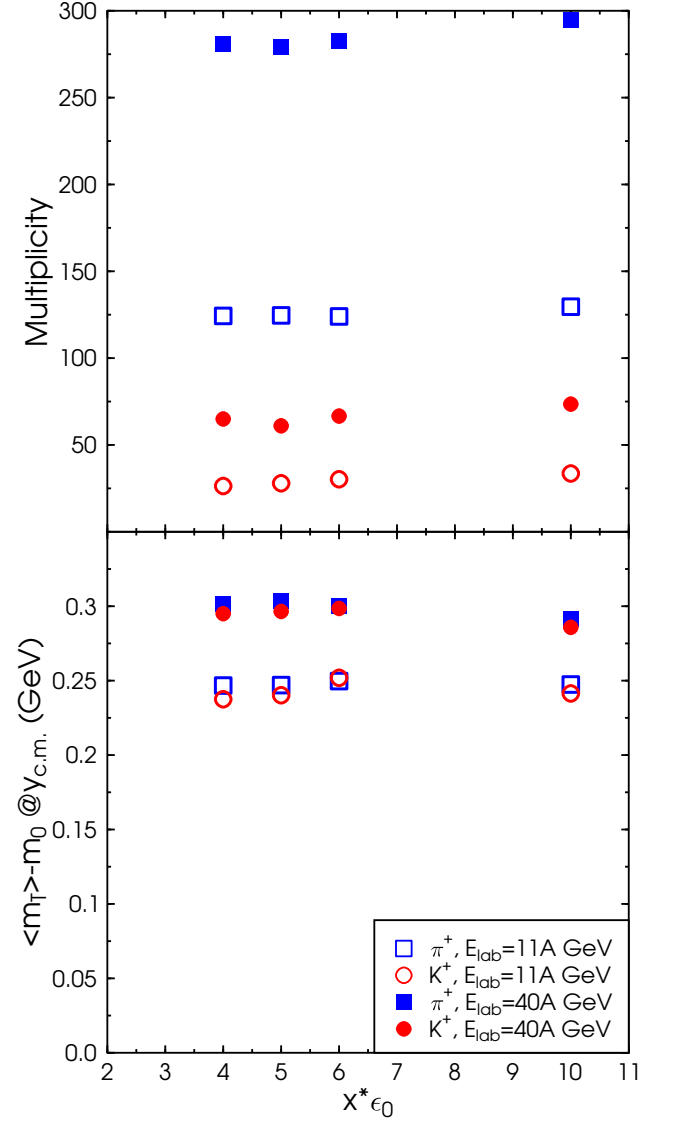}
\caption{\label{fig_frtestmul}
(Color online) Pion and kaon multiplicities (upper panel) and mean transverse mass of pions and kaons at midrapidity ($|y|<0.5$) (lower panel) for four different freeze-out criteria for central ($b<3.4$ fm) Au+Au/Pb+Pb collisions. The open symbols depict the result at $E_{\rm lab}=11A~$GeV while the filled symbols are the results at $E_{\rm lab}=40A~$GeV.}
\end{figure}

The above mentioned steps are pursued on random cells until the initial net baryon number is reached. Strangeness and charge are also conserved in each event separately, energy conservation is fulfilled for the mean values averaged over several events. Aiming at a realistic description of heavy ion reactions we perform the freeze-out for each event separately and do not average over many freeze-outs for one hydrodynamical evolution. 

\section{Parameter tests}

\subsection{Start-time and Freeze-Out Criterium}

In this section we investigate the dependences of observables on parameters of the implementation. Two important parameters have to be determined. The first one is the starting time $t_{start}$ which defines the initial switch from UrQMD to the hydrodynamic evolution. The second parameter is the freeze-out criterium which is parametrized as an energy density criterium. While varying one parameter we have fixed the other one to the default value ($1 \cdot \ t_{start}$ or $5 \cdot \epsilon_0$).

Fig. \ref{fig_intestmul} (top) shows calculations of the total, i.e. pion and kaon ($4 \pi$) multiplicities, for four different starting times at two beam energies. The open symbols depict always the result at the highest AGS energy ($E_{\rm lab}=11A~$GeV) and the filled symbols are the results at the SPS energy ($E_{\rm lab}=40A~$GeV). The starting time is varied in factors of the default value that has been calculated via Eqn. \ref{eqn_tstart}. Displayed are results from halved to doubled initial time. One observes a higher pion production for earlier starting times compared to the pion production in the standard setup (1 $t_{\rm start}$). This may be explained by the fact that the system is forced more strongly to equilibrium and the cascade evolution lasts longer. If the hydrodynamic evolution is started at later times (1.5 or 2 $t_{start}$) the resulting pion multiplicities are not affected anymore. The kaon yield is essentially not sensitive to the switching time. To summarize, varying the starting time by a factor of 4 results in a change in the pion and kaon production of less than $\pm10 \%$ compared to the pion and kaon production in the default configuration (1 $t_{\rm start}$). In Fig. \ref{fig_intestmul} (bottom) the mean transverse mass of pions and kaons at midrapidity is shown. The mean transverse mass values are calculated for the same four different starting times at the two exemplary beam energies as before. Here the results do also not change more than $\pm 15 \%$ for a large spectrum of starting times. Therefore, our choice of the starting time as the geometrical criterium when the nuclei have passed through each other is sensible and stable. It is the earliest possible time where thermalization may have been achieved and the baryon currents have disconnected.  

Fig. \ref{fig_frtestmul} (top) is the equivalent picture to Fig. \ref{fig_intestmul} (top), however displaying the dependence of the total pion and kaon multiplicities on the freeze-out criterium. The default value for the transition energy density is $5 \epsilon_0$ while we have varied it from $(4-10) \epsilon_0$. The higher the freeze-out energy density the earlier the hydrodynamic evolution is stopped because the cells reach this critical energy density value earlier. As a consequence the kaon yields raises with an increase of the energy density criterium, while the pions remain virtually unchanged for all investigated transition criteria. Fig. \ref{fig_frtestmul} (bottom) shows the results for the pions and kaons mean transverse mass for two beam energies and different freeze-out criteria. Again one observes only a very weak dependence on the freeze-out criterion. Furthermore, the mean transverse mass values for the two different meson species are very similar since they acquire the same transverse flow. These findings confirm that our choice for the freeze-out criterium as $5 \epsilon_0$ is robust. 

\subsection{Timescales}

In this Section the time scales that are important will be explored. Let us start with a table that summarizes the mean durations of the hydrodynamic and the hadronic phase of the collision for different starting times and freeze-out parameters at $E_{\rm lab}=40A~$GeV. 

\begin{table}[h]
\begin{center}
\renewcommand{\arraystretch}{1.3}
\begin{tabular}{|c|c|c|c|}
\hline 
$x \cdot t_{start}$ & $ x \cdot \epsilon_0$ & $\langle t_{hydro}\rangle$ [fm]& $\langle t_{hadronic}\rangle$[fm]\\  \hline \hline
1 & 4& 7.68 & 15.63\\ 
1 & 5& 7.72 & 16.07\\
1 & 6& 6.84 & 16.49\\ 
1 &10& 4.60 & 17.29\\
0.5 &5& 7.03 & 14.59\\
1.5 &5& 6.22 & 17.50\\
2 &5& 4.91 & 17.61\\
\hline
\end{tabular}
\caption{\label{tab_params40} This table contains the mean durations of the hydrodynamic evolution and the hadronic calculation afterwards for different starting times and freeze-out criteria for central ($b<3.4$ fm) Pb+Pb collisions at $E_{\rm lab}=40A~$GeV.}
\end{center}
\end{table}

\begin{figure}[t]
\includegraphics[width=0.5\textwidth]{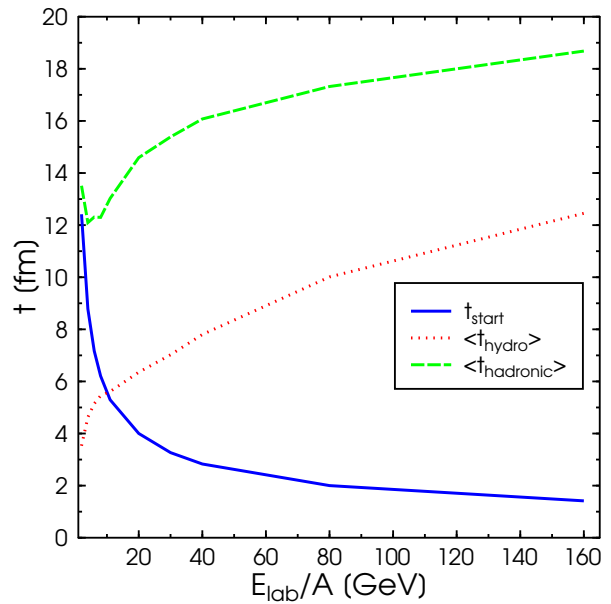}
\caption{\label{fig_timescales}
(Color online) Beam energy dependence of the starting time $t_{start}$ (blue full line), the averaged time for the hydrodynamic evolution $\langle t_{hydro} \rangle$ (red dotted line) and the mean duration of the hadronic stage $\langle t_{hadronic} \rangle$ (green dashed line) of central ($b<3.4$ fm) Au+Au/Pb+Pb collisions.}
\end{figure}

The duration of the hydrodynamic evolution is a well-defined period for each event because of the isochronous freeze-out. The average is therefore an average over 100 events. The hadronic stage starts when the hydrodynamic evolution is stopped and it ends when the particles have undergone their last interaction. An interaction can be an inelastic or an elastic collision or a decay.

The averaged time duration of the stages of the reaction (given in Table \ref{tab_params40}) reflect the expectations. The lower the freeze-out energy density the later the hydrodynamic freeze-out proceeds and therefore the hydrodynamic evolution lasts longer while the hadronic stage is shortened. The later the hydrodynamic evolution starts, the bigger $t_{start}$ is, the shorter the hydrodynamic evolution lasts. The hadronic phase does not show a clear trend. To first approximation the final UrQMD stage lasts for $16.5 \pm 2$ fm independent of the parameters.

Fig. \ref{fig_timescales} shows the beam energy dependences of the timescales for the chosen values of $t_{start}$ and the freeze-out energy density, from $E_{\rm lab} = 2-160A~$ GeV. The starting time decreases as a function of beam energy from more than 10 fm at low energies to less than 2 fm at higher SPS energies. The mean duration of the hydrodynamic as well as the hadronic phase of the reaction grow with raising energy.  

\subsection{Time Evolutions}

\begin{figure}[t]
\includegraphics[width=0.5\textwidth]{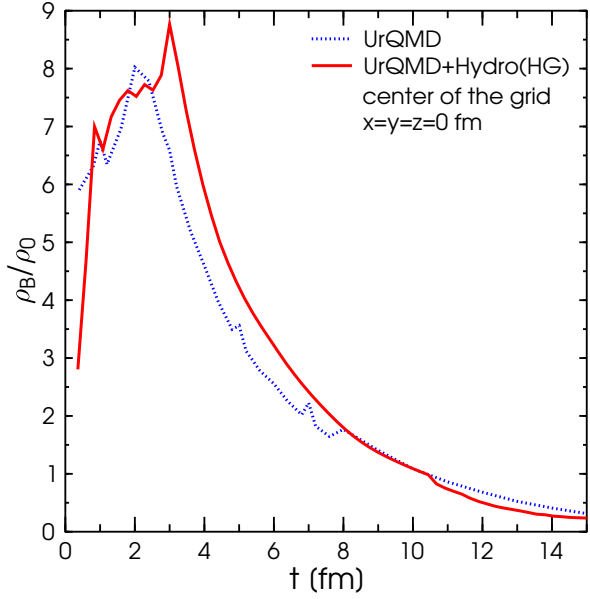}
\caption{\label{fig_rhobtime} (Color online) Time evolution of the net baryon number density in the local rest frame for central ($b=0$ fm) Pb+Pb collisions at $E_{\rm lab}= 40A~$GeV.}
\end{figure}

In the following Section we investigate the time evolution of different quantities and compare the results of the hybrid model calculation to the UrQMD calculation without an hydrodynamic stage. Since the net baryon density is directly propagated on the hydrodynamic grid, it serves as a good example to compare to the default UrQMD calculation. In the microscopic approach the local rest frame density is calculated as the zero component of the net baryon number current in the frame where the corresponding local velocity vanishes \cite{Vogel:2007yu}.

\begin{figure}[t]
\includegraphics[width=0.5\textwidth]{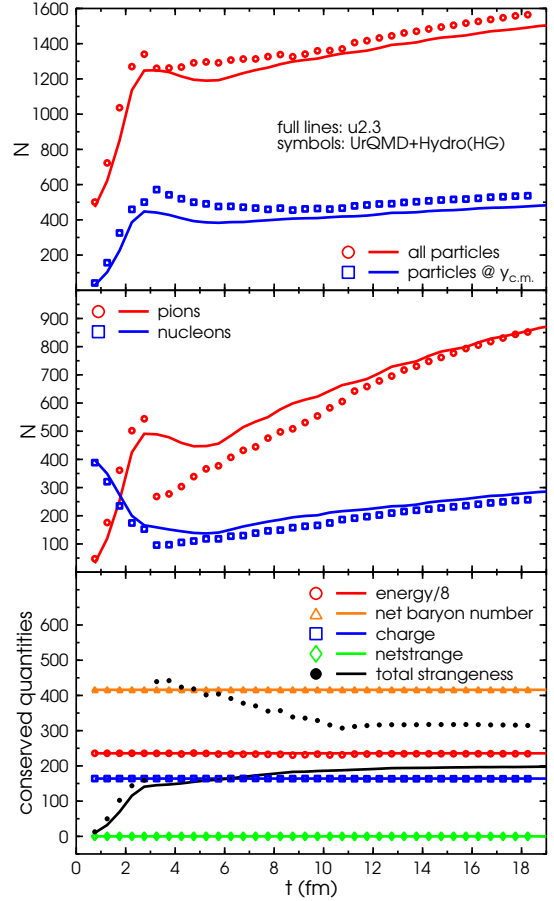}
\caption{\label{fig_noptime} (Color online) Time evolution of the total particle number and the midrapidity ($|y|<0.5$) yield (upper panel), of the total number of pions and nucleons (middle panel) and of the conserved quantities (lower panel) for central ($b=0$ fm) Pb+Pb collisions at $E_{\rm lab}= 40A~$GeV. Results of the hybrid model calculation UrQMD+Hydro (HG) are depicted with symbols while UrQMD-2.3 results are represented by lines. The total energy of the system (red circles and line) has been divided by eight for visibility reasons. The other conserved quantum numbers are net baryon number (orange triangles and line), the overall charge (blue squares and line) and the strangeness (green diamonds and line). The total strangeness (black dots and line) is given by the sum of s- and $\bar{s}$-quarks.}
\end{figure}

Fig. \ref{fig_rhobtime} shows the time evolution of the net baryon number density at the center of a central Pb+Pb collision at $E_{\rm lab}= 40A~$GeV. The blue full line depicts the default UrQMD calculation while the red full line depicts the result of the hybrid model calculation. The result has some spikes because here we compare single events. There are two important observations: The absolute values of the net baryon number densities are very similar in both cases and there are no obvious discontinuities at the switching points to and from the hydrodynamic model calculation. The smoothness of the curve confirms our choice of parameters.

In Fig. \ref{fig_noptime} (top and middle) the time evolution of the particle yields in the two different models for the most central Pb+Pb collisions at $E_{\rm lab}=40A~$GeV  is compared. The multiplicities at different timesteps are extracted from the hydrodynamic evolution by converting the number densities to particle numbers via the freeze-out procedure (see Section \ref{freeze-out}). Fig. \ref{fig_noptime} (top) depicts the total particle multiplicity (red circles and full line) and the midrapidity multiplicity (blue squares and full line). The full lines indicate the default UrQMD calculation, while the symbols show the results of the hybrid model. The multiplicities increase rapidly in the initial 3 fm/c and then decrease a little, followed by a slower constant raise until the final multiplicity is reached. This qualitative behaviour is very similar in both approaches. The decrease of the multiplicity can be associated with the thermalization because absorption and production processes are on the same order. Note again that there are no discontinuities at the switching times in the hybrid calculation.

Next, we explore the time evolution for two particle species in more detail. In Fig \ref{fig_noptime} (middle), the pions (red circles and full line) represent the newly produced particles while the nucleons (blue squares and full line) are already there in the beginning as the constituents of the two incoming nuclei. The qualitative behaviour of the temporal evolution of the pion yield is similar to that discussed above for the total multiplicity. The decrease at the starting time of the hydrodynamic evolution $t \sim 3$ fm/c is much stronger than in the model without hydrodynamic phase, because of the instant thermalization at the transition time. The default UrQMD transport calculation results in a similar, but much smoother, shape of the curve. This similarity hints to the fact that the microscopic calculation also equilibrates the hot and dense matter to a rather large degree. The number of nucleons decreases in the beginning due to the production of resonances and string excitations. At the thermalization the minimum is reached and the number of nucleons increases slowly until the final value is reached. In this case, not only the qualitative behaviour is independent of the underlying dynamics but also the absolute values are very close to each other.

To check quantum number and energy conservation the temporal evolution of all the important quantities in both approaches for central Pb+Pb collisions at $E_{\rm lab}=40 A~$GeV is depicted in Fig. \ref{fig_noptime} (bottom). The default UrQMD calculations are indicated by full lines while the hybrid model calculations with integrated hydrodynamic evolution are depicted by symbols. The net baryon number (orange triangles and full line), the charge (blue squares and full line) and the net strangeness (green diamonds and full line) are exactly conserved in both approaches. The total energy (red circles and full line) is only on average over several events conserved in the hybrid model calculation due to the freeze-out prescription. But the fluctuations are on a $5\%$ level. Note however, that the total strangeness in the system ($s+\bar{s}-$ quarks) is very different in both approaches. In the default transport calculation (black line) the total strangeness increases in the early stage of the collision and remains constant. This is contrasted by the hybrid calculation (dots). Due to the local thermal equilibration and the thermal production of strange particles in the hybrid calculation the yield of strange quarks jumps to a higher value at the switching time ($t_{start}$). The total strangeness then decreases as the system cools down, but the final value remains $50\%$ higher than in the default transport calculation. 

\subsection{Final State Interactions}

\begin{figure}[t]
\vspace{-0.5cm}
\includegraphics[width=0.5\textwidth]{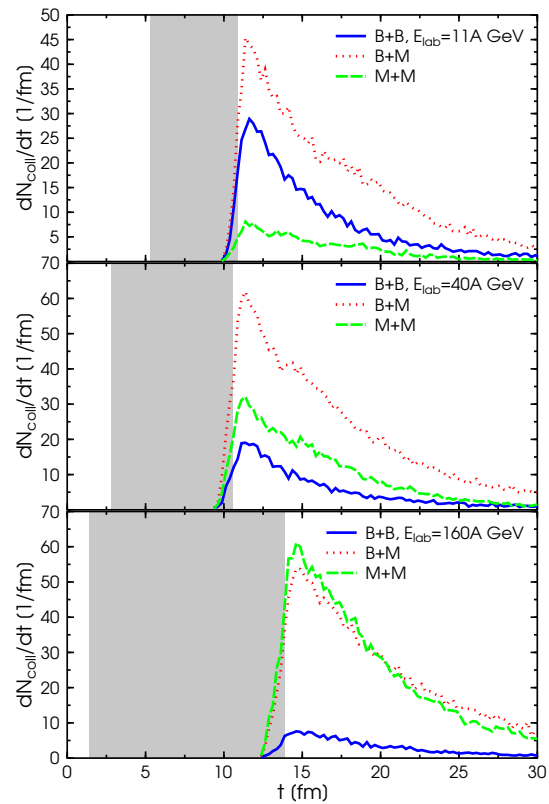}
\caption{(Color online) Temporal distribution of binary collisions in the hadronic cascade calculation after the hydrodynamic evolution. The upper plot depicts the result at $E_{\rm lab}=11A~$GeV, the middle plot at $E_{\rm lab}=40A~$GeV and the lowest plot at the highest SPS energy ($E_{\rm lab}=160A~$GeV) for central ($b<3.4$ fm) Au+Au/Pb+Pb collisions. The full line refers to baryon-baryon collisions (B+B), the dotted line to baryon-meson collisions (B+M) and the dashed line to meson-meson collisions (M+M). The grey shaded area depicts the averaged duration of the hydrodynamic evolution.}
\label{fig_dncolldt}
\end{figure}

\begin{figure}[t]
\includegraphics[width=0.5\textwidth]{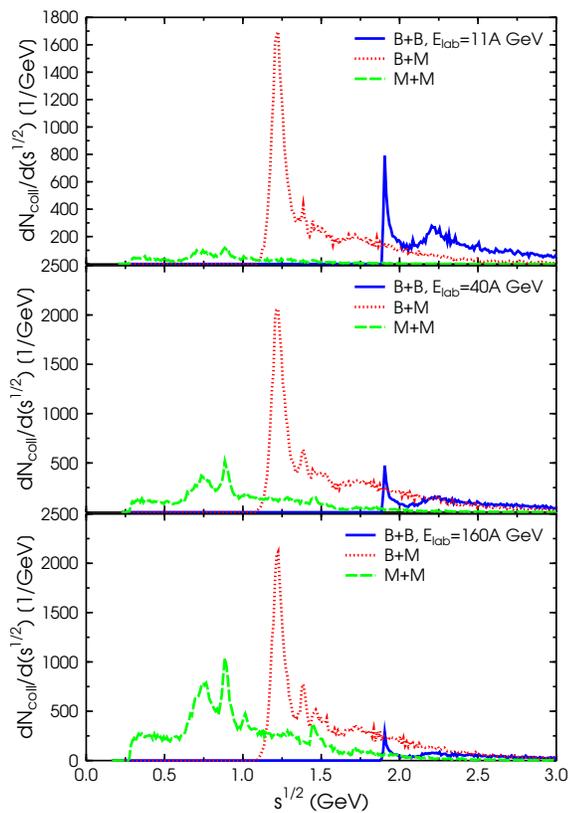}
\caption{(Color online) Distribution of the $\sqrt{s}$ values for the binary collisions in the hadronic cascade calculation after the hydrodynamic evolution. The upper plot depicts the result at $E_{\rm lab}=11A~$GeV, the middle plot at $E_{\rm lab}=40A~$GeV and the lowest plot at the highest SPS energy ($E_{\rm lab}=160A~$GeV) for central ($b<3.4$ fm) Au+Au/Pb+Pb collisions. The full line refers to baryon-baryon collisions (B+B), the dotted line to baryon-meson collisions (B+M) and the dashed line to meson-meson collisions (M+M).}
\label{fig_dncolldsqrts}
\end{figure}

The last step is the analysis of the freeze-out process, i.e. how much hadronic interactions happen after the hydrodynamic evolution. For this purpose, we have calculated the number of collisions during the hadronic cascade calculation in dependence of time and $\sqrt{s}$ of the elementary collisions. The corresponding distributions are shown in Figs. \ref{fig_dncolldt} and \ref{fig_dncolldsqrts}, respectively, for three different beam energies (11, 40\rm{ and }160\rm{$A$ GeV, from top to bottom} ).  

Fig. \ref{fig_dncolldt} shows the collision rates for meson-meson, meson-baryon and baryon-baryon interactions. The grey area indicates the average time span of the hydrodynamic phase. One observes that substantial collision rates are present directly after the transition to UrQMD. The collision rates stay high for 5 fm/c, and after 30-40 fm/c the system is completely frozen out. Only some resonance decays proceed for longer. According to the composition of the system the baryon-baryon and baryon-meson interactions dominate the lower beam energy result, while at the highest SPS energy the meson-meson together with the meson-baryon interactions are the most abundant type of collision indicating the transition from baryon dominated to meson dominated systems. Note that the overlap of the hadronic interaction phase with the hydrodynamic evolution results from the fact that the duration of the different stages fluctuates in the present approach. Shown here is the average duration of the hydrodynamic evolution.

Fig. \ref{fig_dncolldsqrts} shows the $\sqrt{s}$ distribution of the elementary collision in the freeze-out process. One nicely observes all the resonance peaks in the corresponding channels. This figure suggests that the most abundant meson-baryon collisions are excitations of the $\Delta$ resonance (i.e. $\pi N$ interactions) since there is a sharp peak at the $\Delta$ mass ($m_{\Delta}=1.232$ GeV). For meson-meson reactions the $\rho$ and the $\omega$ peaks are clearly present. This result indicates that there is still resonance regeneration even at this late stage of the system's evolution. 

\section{Results}
\subsection{Multiplicities and Particle Spectra}

We start with a comparison of the multiplicities and particle spectra in the two frameworks. Calculations with the embedded hydrodynamic evolution employing a hadron gas equation of state for the high density stage of the collisions are compared to the reference results of the default transport calculation (UrQMD-2.3). Since both calculations use the same initial conditions and freeze-out prescription it allows to extract, which observables are sensitive to the change in the underlying dynamics, thus allowing to explore the effect of local equilibration, of viscosities and heat conductivity. The hybrid model calculation is in the following always depicted as full lines while the default UrQMD-2.3 calculations are depicted as dotted lines. Note that we do not tune any parameters for different energies or centralities for the hybrid model calculation. The starting time for the hydrodynamic expansion is always calculated using Eqn. \ref{eqn_tstart} and the fixed energy density criterium ($5 \epsilon_0$) for the freeze-out (as explained in Section \ref{freeze-out}) is always employed.

\begin{figure}[t]
\includegraphics[width=0.5\textwidth]{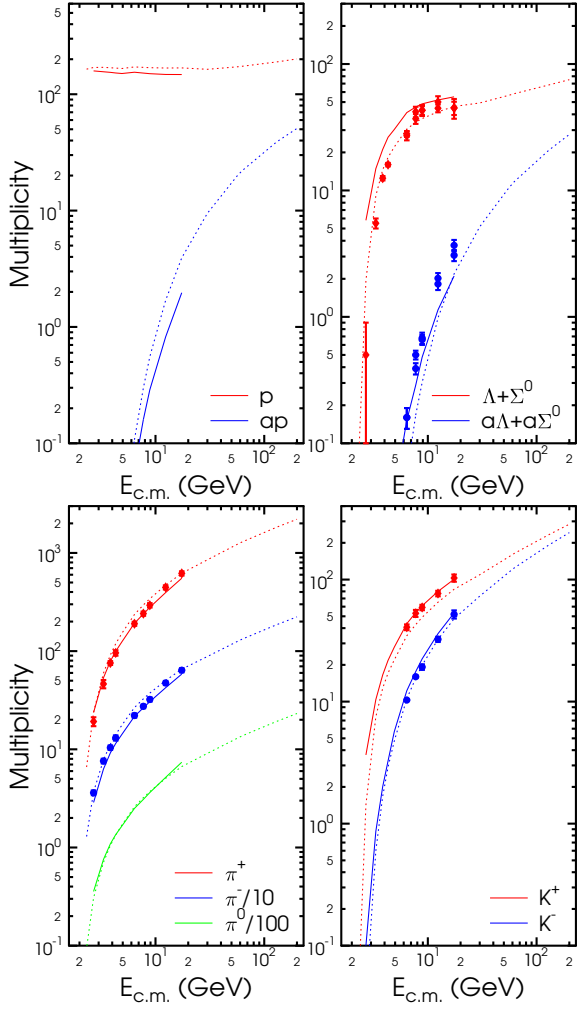}
\caption{(Color online) Excitation function of particle multiplicities ($4\pi$) in Au+Au/Pb+Pb collisions from $E_{\rm lab}=2A~$GeV to $\sqrt{s_{NN}}=200$ GeV. UrQMD+Hydro (HG) calculations are depicted with full lines, while UrQMD-2.3 calculations are depicted with dotted lines. The corresponding data from different experiments \cite{Klay:2003zf,Pinkenburg:2001fj,Chung:2003zr,:2007fe,Afanasiev:2002mx,Anticic:2003ux,Richard:2005rx,Mitrovski:2006js,arXiv:0804.3770,Blume:2004ci,Afanasiev:2002he} are depicted with symbols.}
\label{fig_mulally}
\end{figure}

\begin{figure}[t]
\includegraphics[width=0.5\textwidth]{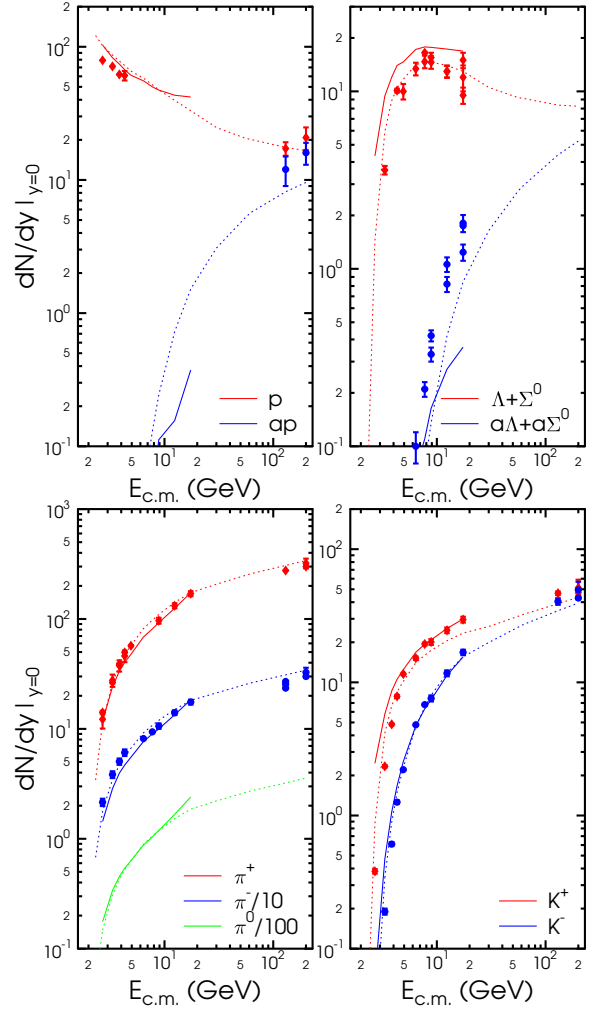}
\caption{(Color online) Excitation function of particle yields at midrapidity ($|y|<0.5$) in Au+Au/Pb+Pb collisions from $E_{\rm lab}=2A~$GeV to $\sqrt{s_{NN}}=200$ GeV. UrQMD+Hydro (HG) calculations are depicted with full lines, while UrQMD-2.3 calculations are depicted with dotted lines. The corresponding data from different experiments \cite{Ahle:1999uy,Klay:2003zf,Afanasiev:2002mx,arXiv:0804.3770,Adcox:2003nr,Lee:2004bx,Ahle:2000wq,Ouerdane:2002gm,Ahmad:1991nv,Mischke:2002wt,Adams:2003xp} are depicted with symbols.}
\label{fig_mulmidy}
\end{figure}

In Fig. \ref{fig_mulally} the excitation functions of the total multiplicities are shown for central Au+Au/Pb+Pb collisions for $E_{\rm lab}=2-160A~$GeV. The present hybrid approach simulations have been restricted to this energy range because for calculations at higher energies some numerical subtleties have to be resolved, e.g. a dynamical grid size for the hydrodynamical evolution. Compared to the default simulation, the pion and proton multiplicities are decreased over the whole energy range in the hybrid model calculation due to the conservation of entropy in the ideal hydrodynamic evolution. The non-equilibrium transport calculation produces entropy and therefore the yields of nonstrange particles are higher. The production of strange particles however, is enhanced due to the establishment of full local thermal equilibrium in the hybrid model calculation. Since the abundance of strange particles is relatively small they survive the interactions in the UrQMD evolution that follows the hydrodynamic freeze-out almost without re-thermalization.

Fig. \ref{fig_mulmidy} shows the midrapidity yields of protons, pions, $\Lambda$'s and kaons as a function of the beam energy for central Au+Au/Pb+Pb collisions at $E_{\rm lab}=2-160A~$GeV. For pions, kaons and $\Lambda$'s the same trend as for the $4\pi$ multiplicities is observed. There are less pions produced in the hybrid model calculation due to entropy conservation, but more strange particles because of the production according to the local thermal equilibrium distributions. The proton yield at midrapidity is very similar in both calculations while there are less antiprotons produced in the hybrid calculation. Also for the strange antiparticles a reduction of the midrapidity yield in the hybrid calculation at higher SPS energies can be seen.  

\begin{figure}[t]
\includegraphics[width=0.5\textwidth]{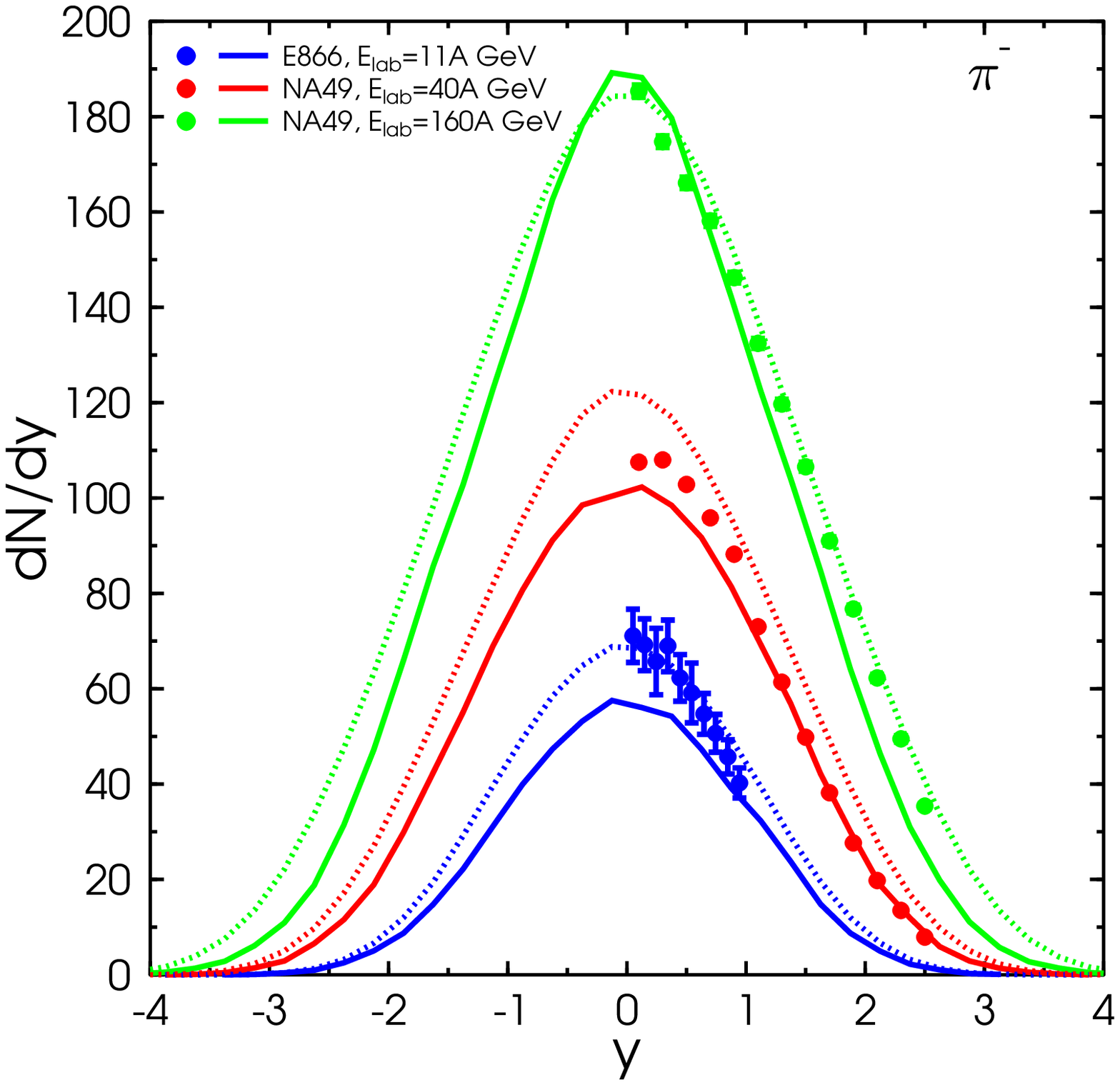}
\caption{(Color online) Rapidity spectra of $\pi^-$ for central ($b<3.4$ fm) Au+Au/Pb+Pb collisions for $E_{\rm lab}=11, 40 \mbox{ and }160A~$GeV. UrQMD+Hydro (HG) calculations are depicted with full lines, while UrQMD-2.3 calculations are depicted with dotted lines. The corresponding data from different experiments \cite{Akiba:1996xf,Afanasiev:2002mx} are depicted with symbols.}
\label{fig_dndypiminus}
\end{figure}

To explore the kinetics of the system in more detail, Fig. \ref{fig_dndypiminus} shows the rapidity distribution for $\pi^-$ at three different energies ($E_{\rm lab} = 11,40$ and $160A~$GeV). The general shape of the distribution is very similar in both approaches and in line with the experimental data. At higher energies even the absolute yields become very close to each other in both approaches. 

\begin{figure}[t]
\includegraphics[width=0.5\textwidth]{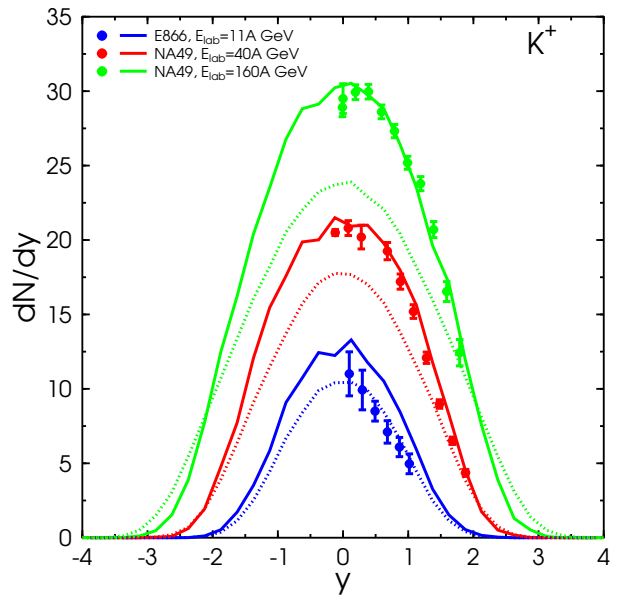}
\caption{(Color online) Rapidity spectra of $K^+$ for central ($b<3.4$ fm) Au+Au/Pb+Pb collisions for $E_{\rm lab}=11, 40 \mbox{ and }160A~$GeV. UrQMD+Hydro (HG) calculations are depicted with full lines, while UrQMD-2.3 calculations are depicted with dotted lines. The corresponding data from different experiments \cite{Akiba:1996xf,Afanasiev:2002mx} are depicted with symbols.}
\label{fig_dndykaplus}
\end{figure}

Fig. \ref{fig_dndykaplus} shows the $K^+$ rapidity distributions. In this case, the yield is higher in the hybrid calculation as already discussed above and also the shape of the distribution fits very nicely to the experimental data at SPS energies. Overall the rapidity distributions seem not to be too sensitive to the details of the dynamics for the hot and dense stage, but strangeness yields are influenced by the local equilibrium assumption. It seems that the local equilibrium assumption provides similar strangeness enhancement as previous calculations including additional strong color fields \cite{Soff:1999et,Bleicher:2000us}. It is remarkable how well the hybrid calculation matches the rapidity spectra at lower energies ($E_{\rm lab}=11A~$GeV), even though the transport calculation provides a slightly better description to the experimental data at this energy. One might still conclude from the rapidity spectra that the local equilibrium is not a good assumption for AGS energies. 

\subsection{Transverse Dynamics}

After the longitudinal dynamics which reflects more the stopping power in the initial state we turn now to the transverse dynamics of the system. Transverse spectra are a promising candidate to be sensitive to the change in the underlying dynamics because they emerge from the transverse expansion which is mostly dominated by the evolution in the hot and dense stage of the reaction.
\begin{figure}[t]
\vspace*{-1cm}
\includegraphics[width=0.50\textwidth]{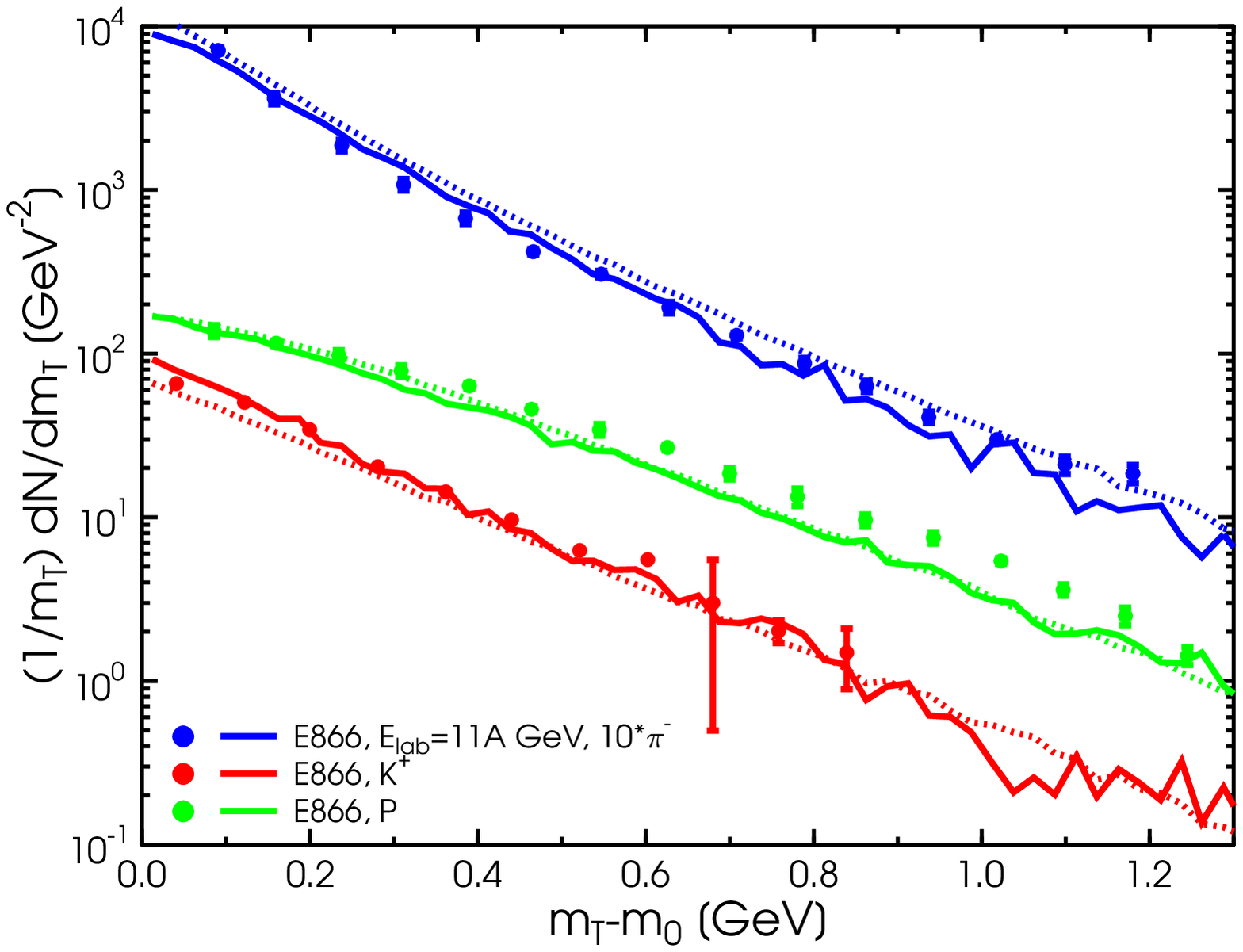}
\caption{(Color online) Transverse mass spectra of $\pi^-$, $K^+$ and protons at midrapidity ($|y|<0.5$) for central ($b<3.4$ fm) Au+Au collisions at $E_{\rm lab}=11A~$GeV. UrQMD+Hydro (HG) calculations are depicted with full lines, while UrQMD-2.3 calculations are depicted with dotted lines. The corresponding data from the E866 experiment \cite{Klay:2003zf,Ahle:2000wq,Akiba:1996xf} are depicted with symbols.}
\label{fig_dndmt11}
\end{figure}
\begin{figure}[t]
\vspace*{-1cm}
\includegraphics[width=0.50\textwidth]{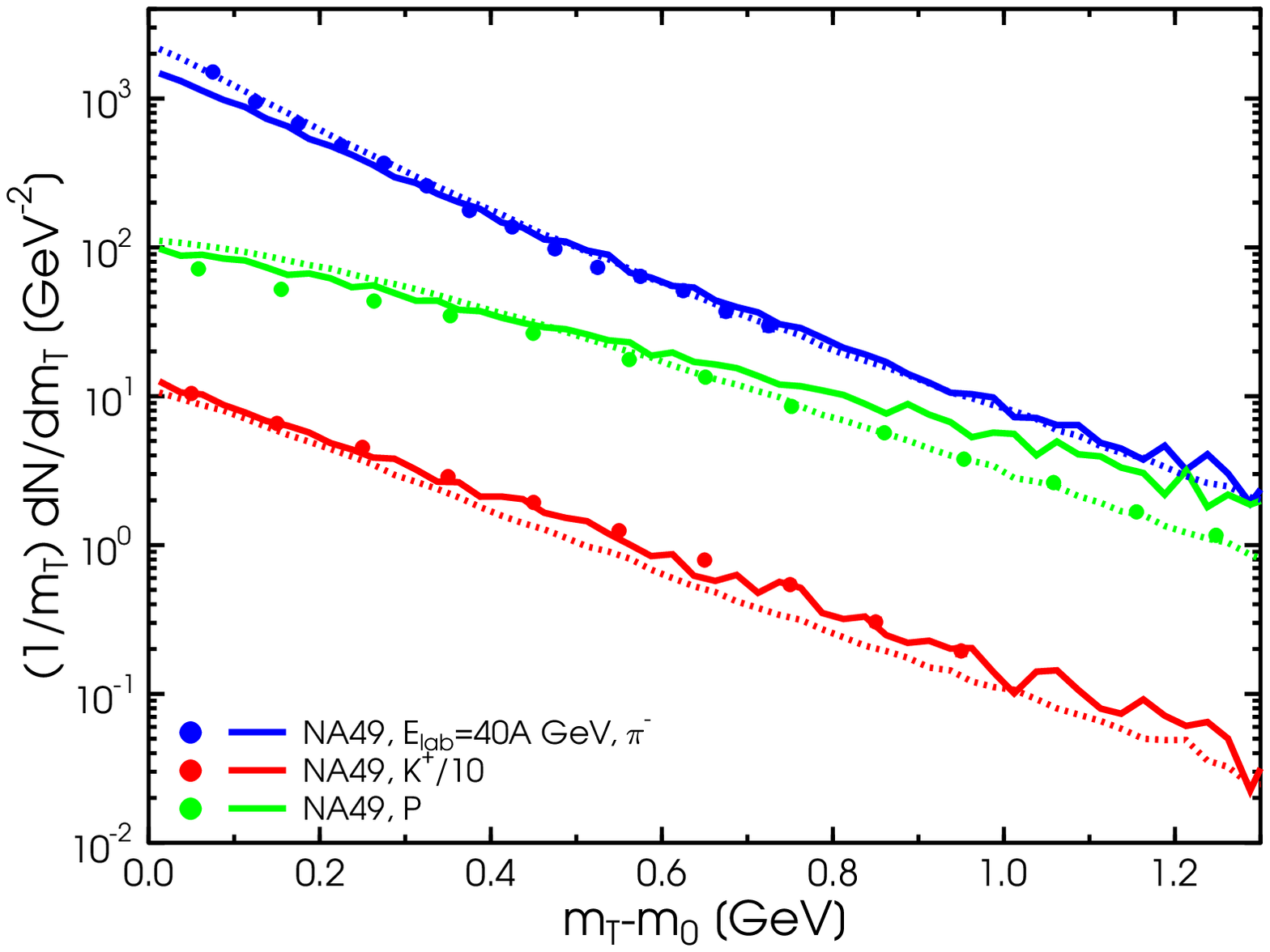}
\caption{(Color online) Transverse mass spectra of $\pi^-$, $K^+$ and protons at midrapidity ($|y|<0.5$) for central ($b<3.4$ fm) Pb+Pb collisions at $E_{\rm lab}=40A~$GeV. UrQMD+Hydro (HG) calculations are depicted with full lines, while UrQMD-2.3 calculations are depicted with dotted lines. The corresponding data from the NA49 experiment \cite{Afanasiev:2002mx,Alt:2006dk} are depicted with symbols.}
\label{fig_dndmt40}
\end{figure}
\begin{figure}[h]
\vspace*{-1cm}
\includegraphics[width=0.50\textwidth]{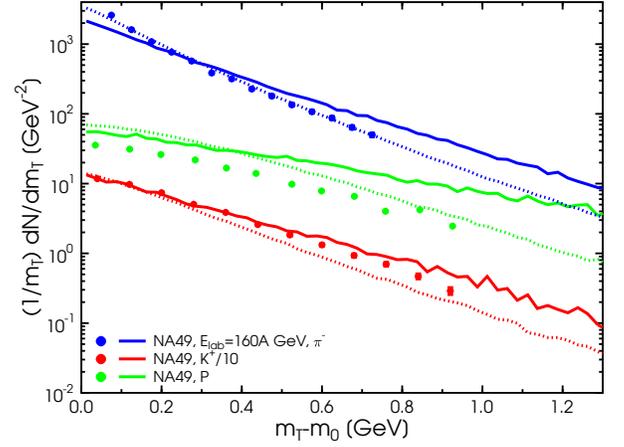}
\caption{(Color online) Transverse mass spectra of $\pi^-$,$K^+$ and protons at midrapidity ($|y|<0.5$) for central ($b<3.4$ fm) Pb+Pb collisions for $E_{\rm lab}=160A~$GeV. UrQMD+Hydro (HG) calculations are depicted with full lines, while UrQMD-2.3 calculations are depicted with dotted lines. The corresponding data from the NA49 experiment \cite{Afanasiev:2002mx,Alt:2006dk} are depicted with symbols.}
\label{fig_dndmt160}
\end{figure}

Figs. \ref{fig_dndmt11}, \ref{fig_dndmt40} and \ref{fig_dndmt160} display the transverse mass spectra for pions, protons and kaons at midrapidity for central Au+Au/Pb+Pb reactions at three different beam energies. At $E_{\rm lab}=11A~$GeV (Fig. \ref{fig_dndmt11}) the differential transverse mass spectra are very similar for both calculations and are in line with the experimental data.

At $E_{\rm lab}=40A~$GeV (Fig. \ref{fig_dndmt40}) first differences become visible. Most notably is the strong flow of protons in the hybrid approach, that results in an overestimate of protons at high transverse momenta.

\begin{figure}[h!]
\includegraphics[width=0.5\textwidth]{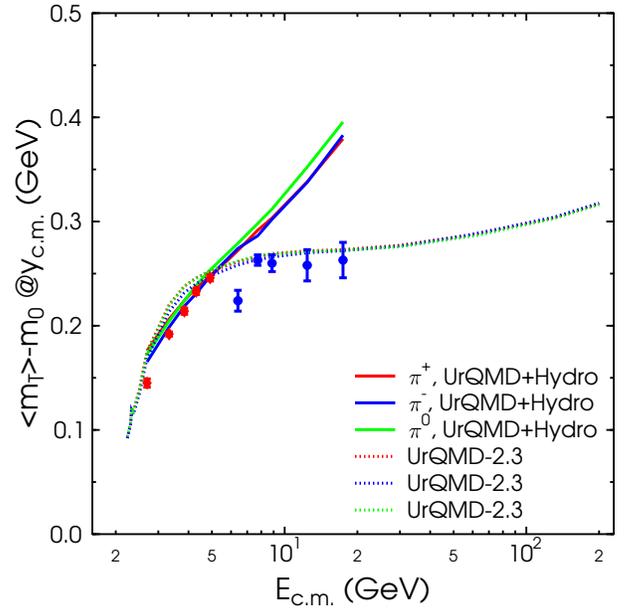}
\caption{(Color online) Mean transverse mass excitation function of pions at midrapidity ($|y|<0.5$) for central ($b<3.4$ fm) Au+Au/Pb+Pb collisions from $E_{\rm lab}=2-160A~$GeV. UrQMD+Hydro (HG) calculations are depicted with full lines, while UrQMD-2.3 calculations are depicted with dotted lines. The corresponding data from different experiments \cite{Ahle:1999uy,:2007fe,Afanasiev:2002mx} are depicted with symbols.}
\label{fig_mmt}
\end{figure}

At the highest SPS energy ($E_{\rm lab}=160A~$GeV, Fig. \ref{fig_dndmt160}) all the transverse mass spectra are flatter in the hybrid approach. The initial pressure gradients are higher in the hydrodynamic calculation due to the hadronic equation of state without phase transition. Therefore, the matter expands faster in the transverse plane and higher transverse masses are reached. At this energy either the introduction of a mixed phase (first order phase transition) or non-equilibrium effects are necessary to explain the experimental data.

\begin{figure}[t]
\includegraphics[width=0.5\textwidth]{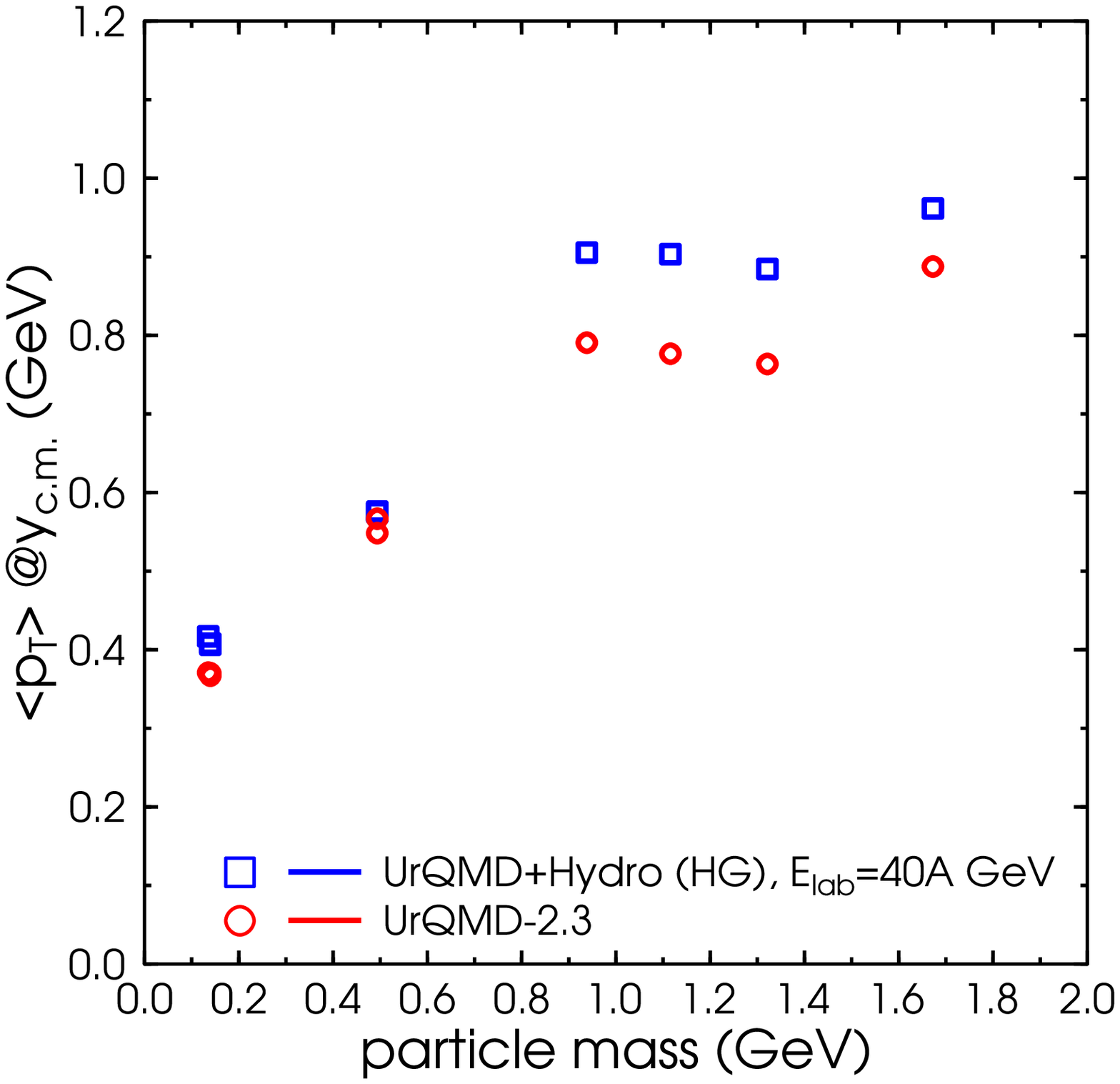}
\caption{(Color online) Mean transverse momentum at midrapidity ($|y|<0.5$) as a function of the particle mass for pions, kaons, protons, $\Lambda$, $\Xi$ and $\Omega$ in central ($b<3.4$ fm) Pb+Pb collisions at $E_{\rm lab}=40A~$GeV. UrQMD+Hydro (HG) calculations are depicted with blue squares, while UrQMD-2.3 calculations are depicted with red circles.}
\label{fig_mptmass}
\end{figure}

Fig. \ref{fig_mmt} shows the mean transverse mass excitation function for pions. It confirms the observations from the differential spectra. Up to $10$ GeV beam energy the hybrid model calculation leads to similar results as the default UrQMD calculation and is in line with the experimental data. The mean value of the transverse mass of pions is proportional to the temperature of the system and very different in the two calculations at higher energies. The UrQMD approach shows a softening of the equation of state in the region where the phase transition is expected because of non-equilibrium effects, while the hadron gas hydrodynamic calculation continuously rises as a function of the energy. This behaviour is well known, see, e.g., \cite{Sorge:1997nv}. 

Finally Fig. \ref{fig_mptmass} shows the mean transverse momenta as a function of particle mass for $\pi, K, p, \Lambda, \Xi,$ and $\Omega$ particles. Here we observe the behaviour known from previous hybrid studies, that with increased strangeness, baryons accumulate less flow than in a complete hydrodynamic approach. This effect can be traced back to the small cross sections of multi strange baryons in the hadronic cascade, thus showing, that the freeze-out/decoupling process proceeds gradually.

\section{Summary}
\label{summary}
We have presented the first fully integrated Boltzmann+hydrodynamics approach to relativistic heavy ion reactions. This hybrid approach is based on the Ultra-relativistic Quantum Molecular Dynamics (UrQMD) transport approach with an intermediate hydrodynamical evolution for the hot and dense stage of the collision. The specific coupling procedure including the initial conditions and the freeze-out prescription have been explained. The event-by-event character of the hybrid approach has been emphasized. The present implementation allows to compare pure microscopic transport calculations with hydrodynamic calculations using exactly the same initial conditions and freeze-out procedure. 

The parameter dependences of the model have been investigated and the time evolution of different quantities were explored. These tests led to the conclusion that the choice of the starting time and the freeze-out criterium does generally alter the multiplicities and transverse mass spectra only on a $20\%$ level. The time evolution has shown that there are no discontinuities at the switching times in the hybrid model calculation. The importance of the final state interactions has been emphasized by demonstrating that there is still resonance regeneration after the hydrodynamic evolution. 

The effects of the change in the underlying dynamics - ideal fluid dynamics vs. non-equilibrium transport theory - have been explored. The final pion and proton multiplicities are lower in the hybrid model calculation due to the isentropic hydrodynamic expansion while the yields for strange particles are enhanced due to the local equilibrium in the hydrodynamic evolution. The results of the different calculations for the mean transverse mass excitation function, rapidity and transverse mass spectra for different particle species at three different beam energies have been discussed in the context of the available data. The transverse expansion of the system is much faster in the hybrid model calculation, especially at higher energies which leads to differences in the observables that are sensitive to the transverse dynamics. This finding indicates qualitatively that ``new'' physical effects like, e.g., non-equilibrium effects or a phase transition have to be taken into account. 

Forthcoming work will be devoted to the study of different equations of state and the effect of changes in the equation of state on observables. Also in progress are calculations within the hybrid model at higher beam energies (RHIC and LHC), but therefore specific numerical subtleties have to be resolved like, e.g., a dynamical grid size for the hydrodynamical evolution.

\begin{acknowledgments}

We are grateful to the Center for Scientific Computing (CSC) at Frankfurt for providing the computing resources. The authors thank Dirk Rischke for providing the 1-fluid hydrodynamics code. H. Petersen gratefully acknowledges financial support by the Deutsche Telekom-Stiftung and support from the Helmholtz Research School on Quark Matter Studies. This work was supported by GSI and BMBF. 
\end{acknowledgments}
%
%\clearpage

\end{document}